\documentclass[aps,prd,12pt,nofootinbib]{revtex4}

\def\blfootnote{\xdef\@thefnmark{}\@footnotetext}

\long\def\symbolfootnote[#1]#2{\begingroup%
\def\thefootnote{\fnsymbol{footnote}}\footnote[#1]{#2}\endgroup}

\newcommand{\be}{\begin{eqnarray}}
\newcommand{\ee}{\end{eqnarray}}
\newcommand{\bcent}{\begin{center}}
\newcommand{\ecent}{\end{center}}
\newcommand{\benum}{\begin{enumerate}}
\newcommand{\eenum}{\end{enumerate}}
\newcommand{\bdesc}{\begin{description}}
\newcommand{\edesc}{\end{description}}
\newcommand{\bitem}{\begin{itemize}}
\newcommand{\eitem}{\end{itemize}}
\newcommand{\bhead}{\begin{center}\bf \Large}
\newcommand{\ehead}{\end{center}\bigskip}
\newcommand{\non}{\nonumber\\}

 \newcommand{\bfk}{{\bf k}} 
 \newcommand{\bfl}{{\bf l}}

 \newcommand{\bfp}{{\bf p}}

 \newcommand{\hatO}{{\hat{O}}}

 \newcommand{\hatrho}{{\hat{\rho}}}

 \newcommand{\hatphi}{{\hat{\phi}}}

 \newcommand{\tildel}{{\tilde{l}}}

 \newcommand{\calK}{{\cal K}}
 
 \newcommand{\calM}{{\cal M}}

 \newcommand{\calS}{{\cal S}}
 \newcommand{\calT}{{\cal T}}

 \newcommand{\bra}[1]{\langle #1 |}
 \newcommand{\ket}[1]{| #1 \rangle}

 \newcommand{\tw}{\textwidth}
 
\usepackage{epsfig}
\usepackage{psfig}
\usepackage{amsmath}	%American Mathematical Society math package
\usepackage{amssymb}    %American Mathematical Society symbol package

\begin{document}

%%%%%%%%%%%%%%%%%%%%%%%%%%%%%%%%%%%%%%%%%%%%%
% Title, authors, affiliations and abstract %
%%%%%%%%%%%%%%%%%%%%%%%%%%%%%%%%%%%%%%%%%%%%%

\pagestyle{empty}

\title{All orders Boltzmann collision term from the multiple scattering
expansion of the self-energy}

%\author
%{Fran\c{c}ois Fillion-Gourdeau, Jean-S\'{e}bastien Gagnon, Sangyong
%Jeon}
%\affiliation
%{Department of Physics, McGill University, 3600 University Street,
%    Montreal, Canada H3A 2T8}

\author{Fran\c{c}ois Fillion-Gourdeau}
\email{ffillion@hep.physics.mcgill.ca}

\author{Jean-S\'{e}bastien Gagnon}
\email{gagnonjs@physics.mcgill.ca}

\author{Sangyong Jeon}
\email{jeon@physics.mcgill.ca}

\affiliation{Department of Physics, McGill University, 3600 University Street, Montreal, Canada H3A 2T8}

\date{\today}

\begin{abstract}
Starting from the Kadanoff-Baym relativistic transport equation and the multiple scattering expansion of the self-energy, we obtain the Boltzmann collision terms for any number of participating particles to all orders in perturbation theory within a quasi-particle approximation.  This work completes a program initiated by Carrington and Mr\'{o}wczy\'{n}ski and developed further by present authors and Weinstock in recent literature.
\end{abstract}

\maketitle

%\pacs{}

\pagestyle{plain}

%%%%%%%%%%%%%%%%%%%%%%%%%%%%%%%
% Main body of the manuscript %
%%%%%%%%%%%%%%%%%%%%%%%%%%%%%%%

%%%%%%%%%%%%%%%%%%%%%%%%%%%%%%%%%%%%%%%%%%%%%%%%%%%%%%%%%%%%%%%%%%%%%%%
\section{Introduction}
\label{Intro}
%%%%%%%%%%%%%%%%%%%%%%%%%%%%%%%%%%%%%%%%%%%%%%%%%%%%%%%%%%%%%%%%%%%%%%%

\setcounter{page}{1}

The study of non-equilibrium phenomena is one of the most interesting
problems
in physics.  In particular, the study of the formation and evolution of
the
quark gluon plasma (QGP) in heavy ion collisions
is both theoretically and experimentally challenging.  The current
dynamical picture of the evolution of the ``fireball'' involves many
stages.
On one hand, it is clear from a theoretical point of view that the
saturated
gluon state (the so-called Color Glass Condensate) describes the
initial
states of the relativistic nuclei
\cite{McLerran_etal_1994_1, McLerran_etal_1994_2, McLerran_2005,
Venugopalan_2005, Kharzeev:2000ph, Kharzeev:2001yq, Kharzeev:2002ei}.
On the other hand, it is
an experimental fact that the final states only consist of fully formed
hadrons. In between these two extremes, the hot and dense matter
evolves
through many stages; after the initial collision, the quarks and
gluons must
scatter enough times to thermalize and form a QGP.
This lump of QGP then must expand
hydrodynamically, eventually reaching low enough temperatures to
undergo a phase transition to hadronic matter.
These hadrons interact further until the final
kinetic freezout temperature is reached.
None of these stages are static nor are they in global equilibrium.

In principle, quantum field theory is capable of explaining all of the
above
processes: Schwinger and Keldysh taught us how to formulate the problem
a long time ago \cite{Schwinger_1961,Keldysh_1964}.  Actually solving a
non-equilibrium problem is, of course, a difficult task. 
For non-equilibrium problems, the most frequently used theoretical
tools are linear response theory, hydrodynamics and kinetic theory.
Linear
response theory and hydrodynamics assume that the system under
consideration is
in a near-equilibrium state.  In fact these two approaches are related in the
sense that
the transport coefficients needed in
hydrodynamics can be calculated using linear response theory
\cite{Kadanoff_Martin_1963}.  For these
two approaches, their formalism is general and does not rely on
perturbation
theory. Although perturbative calculations are almost always the only
practical way to calculate the transport coefficients,
in principle the validity of these approaches
depends {\em only} on the assumption that the system is near equilibrium.

Kinetic theory, in this sense, is different. A kinetic theory is a
theory
of colliding particles and as such, it must rely on the smallness of
coupling
constants
\cite{Jeon_1995,Jeon_Yaffe_1996,AMY_2000,AMY_2003,AMY_2003_2}.  First
of all, to have well-defined particle degrees of
freedom, collision rates must be small enough so that the final state
particles of an individual collision have enough time to become
on-shell (decohere) before being involved in the next collision.
Second, the decay rate of the particles
must be small enough so that the particles become on-shell first and
then decay.
The strong point of kinetic theory is that as long as these
conditions are satisfied, one can numerically study arbitrarily
out-of-equilibrium systems.  The success of the so-called `cascade
models' or
`transport models' for modelling relativistic heavy ions collisions
attests the strength of such an approach.

Current transport models rarely go beyond including $2 \to 2$ and $2
\to 3$
processes  and  resonance decays.  In a recent paper, Carrington and
Mr\'{o}wczy\'{n}ski \cite{Carrington_etal_2005} addressed the
multi-particle
interactions in kinetic theories starting from the Schwinger-Keldysh
formulation of quantum field theory 
(see also Ref.\cite{Mrowczynski:1989bu}).
For a QGP and
also for the early universe,
we expect the systems to be dense enough
for multi-particle interactions to become important.  In
Ref.\cite{Baier_etal_2001} for example, it is argued that the thermalization of the
fireball in their scenario would not be fast enough without $gg
\rightarrow
ggg$ scatterings.  Other studies also stress the
importance of multi-particle interactions (e.g.
\cite{AMY_2003_2,Srivastava_Geiger_1999,Wong_2004,Xu_Greiner_2005}).  In the context of
transport coefficients, it has been shown that for bulk viscosity in
hot scalar
theories, number changing processes are essential and must be included
\cite{Jeon_1995,Jeon_Yaffe_1996}.

The method developped in Ref.\cite{Carrington_etal_2005} is therefore
important
because it provides a systematic way to obtain relativistic transport
equations
that include multi-particle interactions, starting from
first principles.
Using a $g\phi^3+\lambda\phi^4$ scalar field theory, these authors
demonstrated
that up to 4-loop self-energies, the collision term in the
Kadanoff-Baym
equation can be re-expressed as the usual gain and loss rates in a
Boltzmann
equation within a certain quasi-particle ansatz.

As the number of loops increases, the number of relevant diagrams
increases
very quickly.
The authors of Ref.\cite{Carrington_etal_2005} managed this
problem by developing a symbolic manipulation program
to reorganize the diagrams in terms of vacuum scattering processes.
The method
could in principle be applied to any loops and any theories, but at
higher
orders, the task quickly becomes cumbersome.

The goal of this paper is to generalize the results of
\cite{Carrington_etal_2005} using a different method.  Instead of using
a
perturbative expansion, we use the multiple scattering expansion
of the
self-energy described in Ref.\cite{Jeon_Ellis_1998} to show that the
pattern
exhibited in Ref.\cite{Carrington_etal_2005} is in fact general.

A summary of our earlier work on this subject has been published in the
proceedings of Quark Matter 2005 \cite{Gagnon_etal_2005}.  A similar
work was also made
public \cite{Weinstock_2005} shortly after, whose main results 
included only tree-level scattering matrix elements.  The present
contribution completes these works so that the analysis can
be extended to all orders of perturbation theory involving any number
of participating particles.

The rest of the paper is organized as follows.  To set the notations,
and also set the stage for the rest of the paper,
we present in Section \ref{sec:KB} a brief
derivation of the Kadanoff-Baym equation for scalar field theories,
emphasizing the approximations used.  In Section \ref{sec:Expansion},
we
summarize the derivation of the multiple scattering expansion of the
self-energy and introduce our quasi-particle ansatz.
We present the derivation of the Boltzmann collision term
in Section \ref{sec:derivation} and finally conclude in
Section \ref{sec:Conclusion}.

%%%%%%%%%%%%%%%%%%%%%%%%%%%%%%%%%%%%%%%%%%%%%%%%%%%%%%%%%%%%%%%%%%%%%%%
\section{Kadanoff-Baym Equation}
\label{sec:KB}
%%%%%%%%%%%%%%%%%%%%%%%%%%%%%%%%%%%%%%%%%%%%%%%%%%%%%%%%%%%%%%%%%%%%%%%

The goal of this section is to briefly discuss the Kadanoff-Baym
transport
equation for a scalar field theory.  Since there are many excellent
papers and
books on the subject (e.g.
\cite{Carrington_etal_2005,Mrowczynski_1997,Elze_Heinz_1989,
Baym_Kadanoff_1962,Groot_etal_1980}),
we do not attempt another derivation here.  Instead we focus
on the approximations and assumptions used in the derivation and
the interpretations of the equation.
The notations and discussions in this section largely follow
those in Ref.\cite{Carrington_etal_2005} (with a different $i$
convention for propagators, however).  The starting point is the
following Lagrangian density
for a real scalar field:
\begin{eqnarray}
\label{Lagrangian}
{\cal L}(x) & = & \frac{1}{2}\partial^{\mu}\phi \partial_{\mu}\phi
-\frac{1}{2}m^{2}\phi^{2}-\frac{g}{3!}\phi^{3} -
\frac{\lambda}{4!}\phi^{4}
\end{eqnarray}
To discuss the Kadanoff-Baym equation, it is convenient to define the
following
set of two-point Green's functions:
\be
D^{c}_{O}(x,y) & \equiv &
\theta(x_{0}-y_{0})\left<\hat{O}_{H}(x)\hat{O}_{H}(y)\right> +
\theta(y_{0}-x_{0})\left<\hat{O}_{H}(y)\hat{O}_{H}(x)\right> \\
D^{a}_{O}(x,y) & \equiv &
\theta(y_{0}-x_{0})\left<\hat{O}_{H}(x)\hat{O}_{H}(y)\right> +
\theta(x_{0}-y_{0})\left<\hat{O}_{H}(y)\hat{O}_{H}(x)\right> \\
D^{>}_{O}(x, y) & \equiv & \left< \hat{O}_{H}(x)\hat{O}_{H}(y) \right>
\\
D^{<}_{O}(x, y) & \equiv & \left< \hat{O}_{H}(y)\hat{O}_{H}(x) \right>
=
D^>_{O}(y, x) 
\label{eq:D_bigger}
\ee
Here $\hatO_H(x)$ is a Heisenberg picture operator and the average is
defined with respect to the initial density operator $\left< \cdots
\right>
\equiv {\rm Tr}\,\left(\hatrho_0\, \cdots \right)$.  The superscript
$c$ ($a$)
stands for chronological (anti-chronological) ordering.  In the
language
of the Schwinger-Keldysh closed time path formalism, we have $D^{c}_{O}
=
D^{11}_{O}$, $D^{a}_{O} = D^{22}_{O}$, $D^{>}_{O} = D^{21}_{O}$ and
$D^{<}_{O}
= D^{12}_{O}$, where the superscripts~1~(2) refer to the time branch  where
the time
goes from $-\infty$ to $+\infty$ ($+\infty$ to $-\infty$).  It is also
useful
to define the retarded and advanced two-point Green's functions:
\begin{eqnarray}
iD^{\rm ret}_O(x, y) & \equiv & \theta(x^0 - y^0)\, \left<
[\hatO_H(x),\hatO_H(y)] \right> \\
iD^{\rm adv}_O(x, y) & \equiv & \theta(y^0 - x^0)\,\left<
[\hatO_H(y),\hatO_H(x)] \right> = iD^{\rm ret}_O(y,x)
\label{eq:D_adv}
\end{eqnarray}
In this paper, we will use the symbols $G^{<,>,\;...}$ for
propagators and $\Pi^{<,>,\;...}$ for self-energies.

The derivation of the Kadanoff-Baym equation starts with the following
Schwinger-Dyson equation for the contour propagator
\be
\label{eq:Schwinger_Dyson}
G(x,y) & = & G_{0}(x,y) + i\int_C d^4z\int_C
d^4z'\,G_{0}(x,y)\Pi(z,z')G(z',y)
\ee
where the subscript $0$ signifies free field quantities and the time
contour
runs from $-\infty$ to $\infty$ along the real time axis
and then back to $-\infty$ slightly below the real time axis.
Since we want to derive a transport equation, we need to
turn this propagator equation into an equation for the distribution
function
$f(X,\bfp)$.  To do so, one first needs to turn
Eq.(\ref{eq:Schwinger_Dyson})
into an equation for the Wigner transformed $G^{<,>}$.  Since the
Wigner transform $G^{<,>}(X,p)$ is the quantum analog of the
classical distribution function (when it is positive)
\cite{Baym_Kadanoff_1962,Groot_etal_1980}, we can make the following
ansatz for
our distribution function:
\be
\theta(p^0)G^{<}(X,p) & = & \theta(p^0) \rho(X,p) f(X,p)
\ee
where $\rho(X,p)$ is the Wigner transformed spectral density
\be
\rho(X,p) & = & \int d^4u\, e^{ipu}\, \left< [\hatphi(X+u/2),
\hatphi(X-u/2)]
\right>
\ee
To make further progress, it is necessary to make approximations.
More precisely, we require weak inhomogeneity of the system with
respect to the
inverse characteristic momentum (allowing for the gradient expansion) and
the
Compton wavelength (justifying the quasi-particle approximation) in
addition to the weak couplings. 
These approximations are necessary for transport
theory, and we will come back to them in more details in
Section \ref{sec:Approximations}.
Using these approximations and following the steps in
\cite{Carrington_etal_2005}, we get the Kadanoff-Baym equation for a
scalar
theory:
\begin{eqnarray}
\label{Baym_Kadanoff}
E_{p}\left( \frac{\partial}{\partial t} + \mathbf{v} \cdot \nabla
\right)
f(X,\mathbf{p}) + \nabla V(X,\mathbf{p}) \nabla_{p} f(X,\mathbf{p}) & =
&
\theta(p_{0}) C(X,p)
\end{eqnarray}
along with the mass-shell condition:
\begin{eqnarray}
\label{Mass_shell}
p^{2} - m^{2} + V(X,p) & = & 0
\end{eqnarray}
where $V(X,p) = \Pi_{\delta}(X) + \mbox{Re}\; \Pi^{\rm ret}(X,p)$ is the
Vlasov term (corresponding to an effective mass). Here
$\Pi_{\delta}$ is the tadpole self-energy,
$E_{p}$ is the solution of
the mass-shell condition (\ref{Mass_shell}) and $\mathbf{v} \equiv \frac{\partial
E_{p}}{\partial \mathbf{p}}$ is the velocity of propagation.
The collision term is given by:
\begin{eqnarray}
\label{Collision_term}
C(X,p) & = & \frac{1}{2} \left[ \Pi^{<}(X,p) \left( f(X,\mathbf{p}) +1
\right)
- \Pi^{>}(X,p) f(X,\mathbf{p}) \right]
\end{eqnarray}
where $\Pi^{<,>}(X,p)$ are the Wightman self-energies.
These Wightman
functions are ``completely cut'' self-energies and can be
interpreted as gain and loss rates.
Written in this form, the transport equation is similar to the
Boltzmann equation of classical kinetic theory.  The collision term
includes a
gain term with an appropriate 
Bose-Einstein enhancement factor and a loss term \cite{Baym_Kadanoff_1962}.

The derivation of the collision term Eq.(\ref{Collision_term})
in terms of the scattering processes is
the main goal of this paper. The microscopic gain/loss rates
themselves must be computed using the {\em vaccum}
quantum field theory with the in-medium modified mass.
The procedure followed in
Ref.\cite{Carrington_etal_2005}
is to directly analyze multi-loop (up to four)
self-energy diagrams
using a symbolic manipulation program and express them in terms
of physical processes.  In this paper, we use another method based on
the
multiple scattering expansion of the self-energy.

%%%%%%%%%%%%%%%%%%%%%%%%%%%%%%%%%%%%%%%%%%%%%%%%%%%%%%%%%%%%%%%%%%%%%%%
\section{Scattering Expansion of the Self-Energy}
\label{sec:Expansion}
%%%%%%%%%%%%%%%%%%%%%%%%%%%%%%%%%%%%%%%%%%%%%%%%%%%%%%%%%%%%%%%%%%%%%%%

%%%%%%%%%%%%%%%%%%%%%%%%%%%%%%%%%%%%%%%%%%%%%%%%%%%%%%%%%%%%%%%%%%%%%%%
\subsection{Presentation of the Expansion}
\label{sec:Presentation}
%%%%%%%%%%%%%%%%%%%%%%%%%%%%%%%%%%%%%%%%%%%%%%%%%%%%%%%%%%%%%%%%%%%%%%%

The purpose of this section is to present the central ideas involved in
the
multiple scattering expansion of the self-energy.
We will use it in Section~\ref{sec:derivation}
to express the self-energies $\Pi^{<,>}(X,p)$ in terms
of scattering amplitudes so that $C(X,p)$ in Eq.(\ref{Baym_Kadanoff})
can be
interpreted as a kinetic theory collision term.  For general
equilibrium
theories, the scattering expansion
was accomplished in Ref.\cite{Jeon_Ellis_1998}. Here we
repeat the central points and explain how it should be modified to deal
with
subtleties associated with medium complications.

In this section, we use equilibrium quantum field theory to derive the
expansion.  However, as long as we are careful and do not use the KMS
condition, the results in this section can be readily generalized to
the non-equilibrium case by a simple ansatz $n(E_{p})\to f(X, \bfp)$,
where $f(X, \bfp)$ is the non-equilibrium phase space density.  
This is a very simple extension of the equilibrium
theory to the non-equilibrium case; but as long as the Kadanoff-Baym
equation is valid, it captures a large amount of physics.

The starting point of our derivation is the four
propagators\footnote{In this
paper, we concentrate on scalar theories because we want to reproduce
the
results of \cite{Carrington_etal_2005}. However, the multiple
scattering
expansion is general and does not rely on any specific Lagrangian.}
corresponding to the Green functions in the Schwinger-Keldysh contour.
Denoting the regular $(-\infty,\infty)$ time branch as ``1''  and the
reverse $(\infty, -\infty)$ branch as ``2'' , the free propagators for
equilibrium situations are given by
\cite{Jeon_Ellis_1998,Kobes_Semenoff_1985}:
\be
\label{eq:G11}
G_{11}^0(p) & = & G_{ F}^0(p) + \Gamma^0(p) \\
\label{eq:G22}
G_{22}^0(p) & = & G_{ F}^0(p)^* + \Gamma^0(p) \\
\label{eq:G12}
G_{12}^0(p) & = & D_{ F}^0(p) + \Gamma^0(p) \\
\label{eq:G21}
G_{21}^0(p) & = & D_{ F}^0(-p) + \Gamma^0(p)
\ee
where
\be
G_{ F}^0(p) &=& \frac{i}{p^2 - m^2 + i\epsilon}
\\
D_{ F}^0(p) &=& \theta(p^0)2\pi\delta(p^2 - m^2)
\ee
are the vacuum Feynman propagator and the vacuum Cutkosky propagator,
respectively.  To compute the Wightman self-energy using this set of propagators, one must use the following Feynman rules \cite{Jeon_Ellis_1998}:
\begin{enumerate}
\item Draw all cut diagrams where the cut separates the two external vertices.  Label $\Pi^{>}(k)$ all diagrams where the external momentum $k$ enters the unshaded (`1') region.  Label $\Pi^{<}(k)$ all diagrams where $k$ enters the shaded (`2') region.

\item Use the usual Feynman rules for the unshaded side assigning $G_{11}^0(p)$
to the uncut lines. For the shaded side, use the conjugate Feynman rules assigning
$G_{22}^0(p)$ to the uncut lines.

\item If the momentum of a cut line crosses from the unshaded to the shaded region, 
assign $G_{12}^0(p)$. If the momentum of a cut line crosses from the shaded to the unshaded region, 
assign $G_{21}^0(p)$.

\item Divide by the appropriate symmetry factor.
\end{enumerate}
To see an explicit example employing the above rules, see Appendix \ref{app:2to2}.

Note that each propagator can be separated into a zero
temperature part and a finite temperature part.  Furthermore, the
thermal phase space factor
\begin{eqnarray}
\label{Thermal_phase_space_factor}
\Gamma^0(p) & = & n(|p^{0}|)2\pi\delta(p^{2}-m^{2})
\end{eqnarray}
is {\em common} to all $G_{ij}^0$ and the zero-temperature part of the
propagators are exactly those of the vacuum Cutkosky rules
\cite{Diagrammar}. Therefore any diagram in this approach can be
naturally divided into a thermal part ($\Gamma^0(p)$) and a
zero-temperature part.

Using the above two properties of the finite temperature propagators,
it is possible to expand any diagram as a power series in the number of
thermal phase space factor.  In this expansion, each coefficient of the
expansion is a zero temperature scattering diagram \cite{Jeon_Ellis_1998}.

At a first glance, this looks like a fairly complicated expansion.
First, for any given topology of a diagram, one needs to consider all
relevant thermal cuts (i.e. sum of all different combinations of 1 and
2 vertices) \cite{Kobes_Semenoff_1985}.  Then each of these thermal cut
diagrams must be expanded again according to the number of
$\Gamma^0$'s.  This expansion becomes manageable only if one can
show that the coefficients of the same $\Gamma^0$ factors have a simple
meaning in the context of the vacuum theory.

\begin{figure}[t]
\begin{center}
\includegraphics[width=0.7\tw]{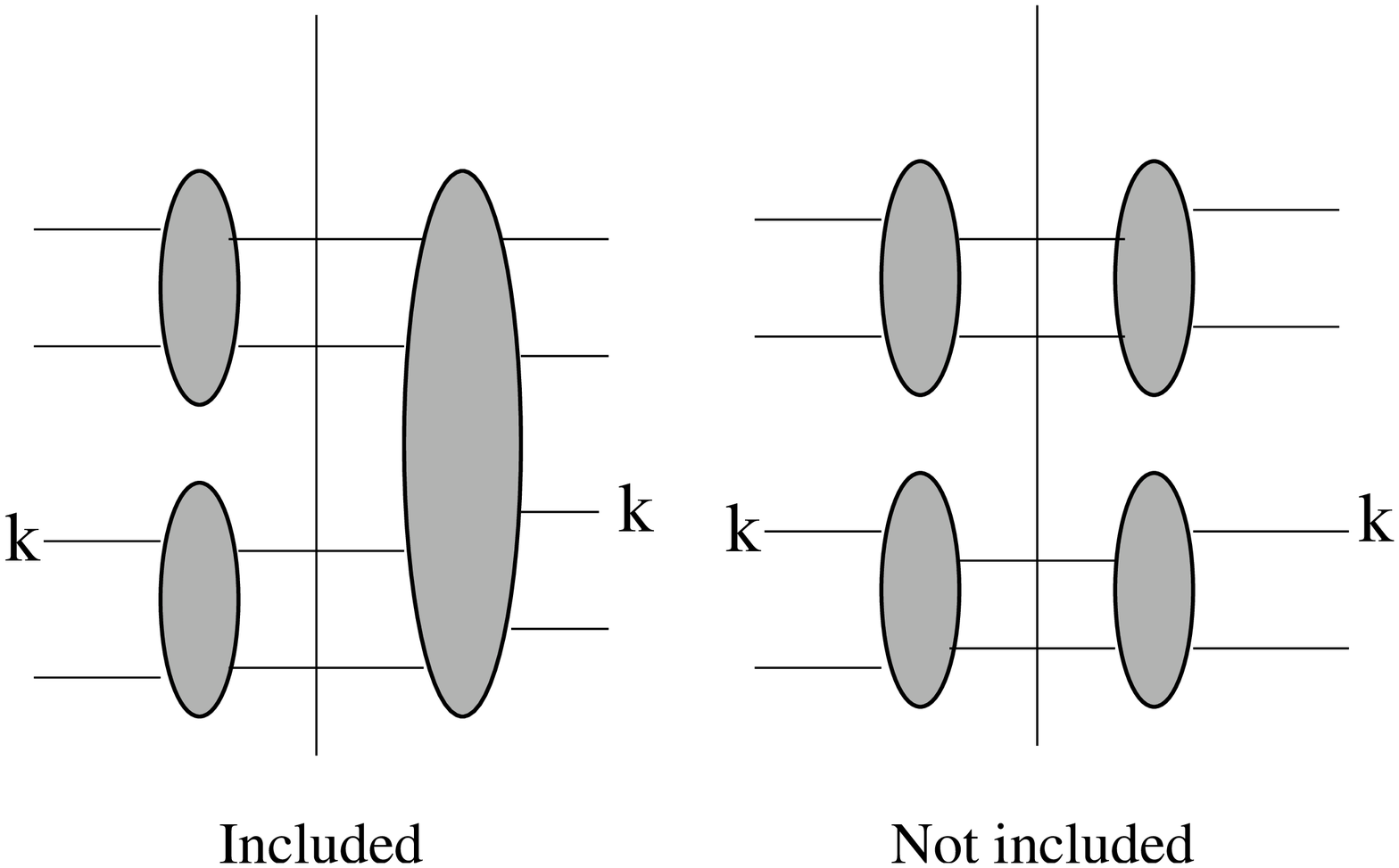}
\end{center}
\caption{
An illustration of diagrams that are included and not included in
$\bra{k,\{l_i\}_m}\calT_k^\dagger\ket{\{p_j\}_n}
 \bra{\{p_j\}_n}\calT_k\ket{k,\{l_i\}_m}
 \Big|_{\rm conn.}$
}
\label{fig:not_include2}
\end{figure}

\begin{figure}[t]
\begin{center}
\includegraphics[width=0.8\tw]{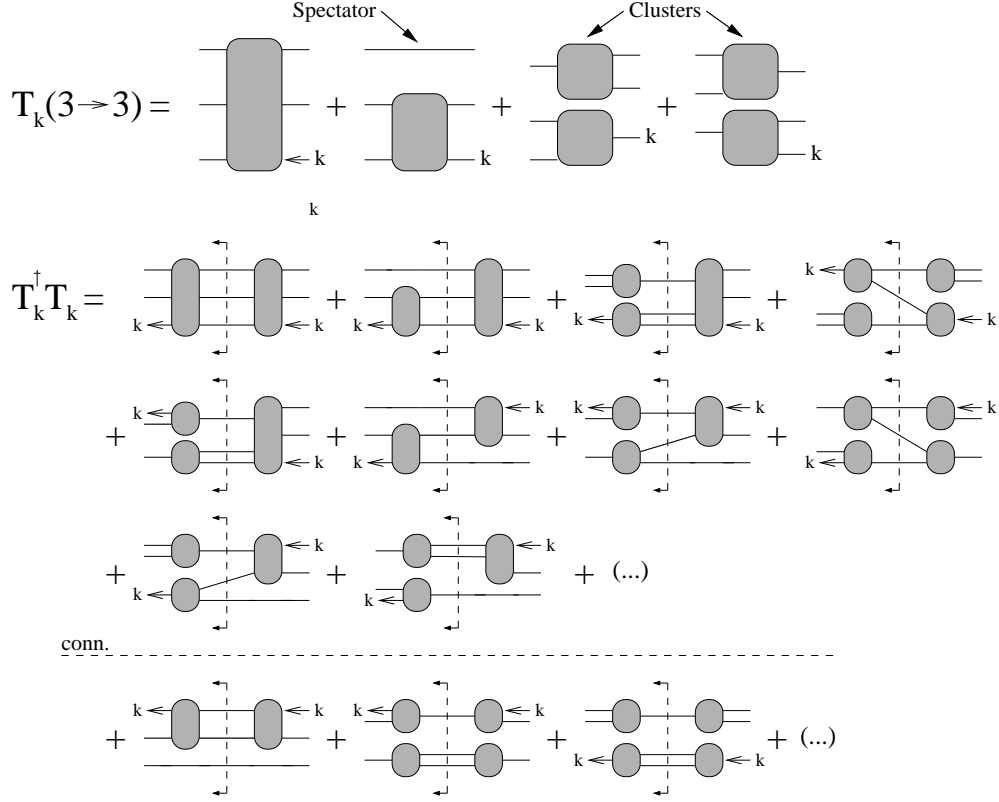}
\end{center}
\caption{An example of a $3 \rightarrow 3$ process included in the multiple scattering expansion $\bra{k,\{l_i\}_2}{\calT_k^\dagger}\ket{\{p_j\}_3} \bra{\{p_j\}_3}{\calT_k}\ket{k,\{l_i\}_2} \Big|_{\rm conn.}$.  The lines cut by the dashed lines are final states and the arrows indicate the complex conjugated side.  The blobs represent connected amplitudes.  The $\{p_j\}_3$ are the momenta of the final states while the $\{l_i\}_2$ are the initial states weighted by thermal factors.  The external momentum $k$ is always interacting. $\calT_k$ contains both clusters and spectators. The expression $\calT_k^\dagger \calT_k$ contains all the possible connections between each other. The $(...)$ represents diagrams similar to the ones already shown but with all momenta and the shading reversed. The subscript \textit{conn.} selects all the diagrams above the dashed line.  The first diagram of $\calT_k^\dagger \calT_k$ is the only \textit{fully connected} diagram.}
\label{fig:T3}
\end{figure}

 That these coefficients cannot be the usual Feynman diagrams is
 evident from the appearance of the cutting rule propagators
 in Eqs.(\ref{eq:G11}-\ref{eq:G21}). Hence,
 the coefficients must correspond to the vacuum Cutkosky (cut)
diagrams.
 In vacuum theory, Cutkosky diagrams are needed when one needs to work
with
 the unitarity condition
 \be
 2\hbox{Im}\,\calT = \calT^\dagger \calT
 \label{eq:unitarity}
 \ee
 where $i\calT = \calS - 1$ is the transition operator.
 Therefore, roughly speaking
 \benum
  \item[(i)] The $G_{ F}^0$ part of
  $G_{11}^0$ propagators form Feynman diagrams
    that correspond to the matrix elements of $\calT$, whereas the
$G_{ F}^0{}^*$ part
    of $G_{22}^0$ propagators form Feynman diagrams
    that correspond to the matrix elements of $\calT^\dagger$.
  \item[(ii)] The $D^0_{ F}$ part
  of $G_{12}^0$ and $G_{21}^0$ propagators provide
  the intermediate states connecting $\calT^\dagger$ and $\calT$ matrix
  elements.
  \item[(iii)] The thermal part $\Gamma^0$ represents
  the bath of {\em on-shell}
  particles in the medium that
  participate in the collision process described
  by the matrix elements of $\calT$ and $\calT^\dagger$.
 \eenum
 In Ref.\cite{Jeon_Ellis_1998}, it was argued that following
 this reasoning yields
 \be
 \Pi^>(k)
 &=&
 \sum_{m=0}^\infty
 \sum_{n=0}^\infty
 {1\over m! n!}
 \left(\int \prod_{i=1}^m {d^3l_i\over (2\pi)^3 2E_{l_i}}\,
n(E_{l_i})\, \right)
 \left(\int \prod_{j=1}^n {d^3p_j\over (2\pi)^3 2E_{p_j}}\, \right)
 \non & & {} \qquad\qquad\qquad\qquad \times
 \bra{k,\{l_i\}_m}{\calT_k^\dagger}\ket{\{p_j\}_n}
 \bra{\{p_j\}_n}{\calT_k}\ket{k,\{l_i\}_m}
 \Big|_{\rm conn.}
 \label{eq:TdaggerT}
 \ee
The subscripts $m$ and $n$ in $\{l_i\}_m$ and $\{p_j\}_n$ indicate that there are $m$ $l_i$'s and $n$ $p_j$'s in each set.  In here and the following, we use the label $k$ exclusively for the external momentum entering the unshaded side of the self-energy diagram.  The subscript $k$ on the operator ${\cal T}_{k}$ means that the external momentum always interacts with the other momentum states.  Note that we do not write $|\bra{\{p_j\}_n}{\calT_k}\ket{k,\{l_i\}_m}|^{2}$  but leave $\bra{k,\{l_i\}_m}{\calT_k^\dagger}\ket{\{p_j\}_n}$ and $\bra{\{p_j\}_n}{\calT_k}\ket{k,\{l_i\}_m}$ explicitly separated; this is to express that even if the momenta $\{p_{j}\}$ can be interconnected in all possible ways, some diagrams needed to make a squared matrix element are missing.  The subscript ``conn.'' indicates that the joining of $\calT_{k}^\dagger$ and $\calT_{k}$ does not give disconnected pieces, even if by itself $\calT_{k}$ contains both clusters and spectators (see Figs. \ref{fig:not_include2} and \ref{fig:T3}). A cluster is a connected amplitude for a certain subset of initial and final momentum states that is completely disconnected from the other amplitudes.  Note here that the denomination ``initial'' and ``final'' when applied to states sandwiching ${\cal T}_{k}$ is slightly different from states sandwiching the fully connected scattering amplitude (see Fig. \ref{fig:Mul_ex} and the discussion around Eqs.(\ref{eq:G12r})-(\ref{eq:G21r})).  A spectator is a state that does not interact with other states (see Fig. \ref{fig:T3}).  For future reference, we also define ``fully connected'' as the sum of all diagrams where all momentum states are connected together when joining \textit{two} amplitudes (see Fig. \ref{fig:T3} for an illustration of this).

%%%%%%%%%%%%%%%%%%%%%%%%%%%%%%%%%%%%%%%%%%%%%%%%%%%%%%%%%%%%%%%%%%%%%%%
\subsection{Medium Related Complications}
\label{sec:Medium}
%%%%%%%%%%%%%%%%%%%%%%%%%%%%%%%%%%%%%%%%%%%%%%%%%%%%%%%%%%%%%%%%%%%%%%%

To turn the collision term of the Kadanoff-Baym equation into the usual
Boltzmann equation collision term,
we must determine under what conditions Eq.(\ref{eq:TdaggerT}) can be approximated by
\be
\Pi^{>}_{\rm Boltz}(k)
& = &
\sum_{s=0}^{\infty} \sum_{n=0}^{\infty} \frac{1}{n!\,s!}
\left(\int\prod_{i=1}^{n}\frac{d^3 l_i}{(2\pi)^3 2E_i} n(E_i)\right)
\left(\int\prod_{j=1}^{s}\frac{d^3 p_j}{(2\pi)^3 2E_j}
[1+n(E_j)]\right)
\non & & {} \qquad\qquad\qquad\qquad \times\big| \langle \{ p_j \}_s |
{\cal M} |k, \{l_i\}_n \rangle \big|^2
\label{eq:PiBoltz}
\ee
where
$
\big| \langle \{ p_j \}_s | {\cal M} |k, \{l_i\}_n \rangle \big|^2
$
is the square of the fully connected scattering matrix element between
the
initial momenta $k$ and $\{l_i\}_n$ and the final momenta $\{p_j\}_s$.
This is the main goal of our paper.  Note that ${\cal M}$ does not carry the subscript $k$ since it is fully connected.

The presence of the medium, however, complicates this task.
Due to the processes occuring within the medium, the
$\bra{\{p_j\}_n}{\calT^{\vphantom{\dagger}}_k}\ket{\{l_i\}_m}$ piece
appearing in Eq.(\ref{eq:TdaggerT})
does not have to be composed of only the fully connected diagrams.
Some of the initial
momenta $\{l_i\}_m$ and the final momenta $\{p_j\}_n$ can interact
among
themselves without interacting with the main piece where $k$
interacts.
This is illustrated in Fig. \ref{fig:not_include2} and Fig. \ref{fig:T3}.  Figure \ref{fig:not_include2} also illustrates why the factor
\be
\calK_T(k,\{l_i\}_{m},\{p_j\}_n)
\equiv
\bra{k,\{l_i\}_m}\calT_k^\dagger\ket{\{p_j\}_n}
 \bra{\{p_j\}_n}\calT_k\ket{k,\{l_i\}_m}
 \Big|_{\rm conn.}
\ee
cannot be readily interpreted as the square of a fully
connected scattering matrix element.
The right hand side diagram in Fig.~\ref{fig:not_include2} 
must appear in the square of the $\calT$ matrix element,
yet it is missing in $\calK_T$ because these types of diagrams
do not appear naturally in the self-energy calculation. This fact is taken into account by the ``conn.'' prescription.

The occurence of `spectators' and `clusters'
in a $\calT$ matrix element is well known
(for instance, see \cite{Weinberg}).
In vacuum theory, these disconnected pieces do not contribute to
the
self-energy calculation because they represent {\em off-shell} vacuum
fluctuations which are independent of the scattering
process.
At finite temperature, however, these in-medium
processes can no longer be neglected.
This is because
the disconnected pieces can now represent real processes occuring
between
on-shell particles in the medium. Furthermore, 
quantum interference can occur between these $k$-independent
in-medium processes and the process where $k$ is directly involved.
The first diagram in Fig.~\ref{fig:not_include2} is exactly of this
type.

Another complication is the fact that no
momentum
state is stable at finite temperature since, within a mean free path,
the momentum of any particle will likely change.
This implies that there are no real asymptotic states at finite
temperature \cite{Landsman_VanWeert_1987,Landsman:1988ta}.
Therefore, one must always
start with resummed propagators and make an appropriate
quasi-particle approximation.

The use of resummed propagators is also essential to deal with
so-called ``pinch'' singularities.  When computing diagrams in finite
temperature quantum field theory, one must be careful with self-energy
insertions since the resulting product of contour propagators 
can give rise to ill defined products of delta
functions.
Fortunately, for equilibrium systems, 
these singularities cancel due to the KMS condition
\cite{Landsman_VanWeert_1987,Altherr_1995,Jeon_Ellis_1998}.  However,
for out-of-equilibrium systems in which the KMS condition is not
satisfied, those singularities do not cancel
\cite{Altherr_Seibert_1994,Altherr_1995,Bedaque_1995,
Greiner_Leupold_1998,Greiner_Leupold_1999} at any finite order in
perturbation theory.  
But as pointed out in \cite{Bedaque_1995,Greiner_Leupold_1998,
Greiner_Leupold_1999}, an out-of-equilibrium system is 
integrated only over a finite volume; thus, 
products of delta functions do not diverge in physical situations. 
As discussed in Refs. \cite{Altherr_1995,Bedaque_1995,
Greiner_Leupold_1998,Greiner_Leupold_1999}, 
the way around this difficulty is to use resummed propagators, 
where no self-energy insertions (and thus no pinch singularities) appear.  To show that Eq.(\ref{eq:PiBoltz}) is indeed a good approximation of Eq.(\ref{eq:TdaggerT}) and to deal with pinch singularities correctly, it is thus essential to use resummed propagators.

%%%%%%%%%%%%%%%%%%%%%%%%%%%%%%%%%%%%%%%%%%%%%%%%%%%%%%%%%%%%%%%%%%%%%%%
\subsection{Resummed Perturbation Theory and the Skeleton Expansion}
\label{sec:Resummed}
%%%%%%%%%%%%%%%%%%%%%%%%%%%%%%%%%%%%%%%%%%%%%%%%%%%%%%%%%%%%%%%%%%%%%%%

The spectral
density for a stable particle in vacuum has two separate
pieces, an on-shell $\delta$-function piece and an off-shell
continuum piece.  Stability requires that these two pieces are well
separated.  In a medium, however, on-shell states are not stable since
elastic collisions can easily change their momenta.
This implies that the
zero temperature part and the in-medium part may not separate cleanly. 

Unless the external particles are either stable or we assume that the
interactions are turned
off at $t = \pm\infty$,
the scattering matrix elements between {\em on-shell} particles
are not well defined.
Technically speaking, the validity of the LSZ
formalism depends on the fact that we can cleanly amputate the external
legs
of a Feynman diagram.  That is, the self-energy insertions along any
external leg result only in renormalization of
the mass and the wavefunction.
At finite temperature, this can no longer be cleanly achieved because
the self-energy has a non-zero imaginary part for any momentum.
In other words, both the on-shell piece and the continuum piece in the
spectral density broaden so that
there is no longer a gap between them.
Nevertheless, if the interaction strength is weak and the medium is
weakly inhomogeneous, the on-shell piece and
the continuum piece should be separated enough so that we can 
use a `quasi-particle' approximation.  We will do so in a short while.
Here, we first discuss the skeleton expansion whose result is needed
later.

To address the self-energy problem in a systematic way,
consider the resummed perturbation
theory where the propagators are fully resummed.
In such a theory, the propagators in Eqs.(\ref{eq:G11}-\ref{eq:G21})
become:
\be
\label{eq:G11_full}
G_{11}(p) & = &
G_F(p) + \Gamma(p) \\
\label{eq:G22_full}
G_{22}(p) & = &
G_F^*(p) + \Gamma(p) \\
\label{eq:G12_full}
G_{12}(p) & = & \theta(p^0)\rho(p) + \Gamma(p) \\
\label{eq:G21_full}
G_{21}(p) & = & -\theta(-p^0)\rho(p) + \Gamma(p)
\ee
where
\be
G_F(p) = \frac{i}{p^{2}-m^{2}+\Pi_{R}(p)+i |\Pi_{I}(p)|}
\ee
is the resummed analogue of the Feynman propagator
and
\begin{eqnarray}
\label{eq:Spectral_density}
\rho(p) = \frac{2\Pi_{I}(p)}{|p^{2} - m^{2} + \Pi_{R}(p) +
i\Pi_{I}(p)|^{2}}
\end{eqnarray}
is the resummed spectral density.  The thermal factor is given by:
\be
\Gamma(p) = n(|p^0|)\mbox{sgn}(p^0)\rho(p)
\ee
In these expressions,
$\Pi_{R}(p)$ and $\Pi_{I}(p)$ are the real and imaginary parts
of the {\em retarded} self-energy and
$\hbox{sgn}(x)$ denotes the sign of $x$.

The original multiple scattering expansion of the self-energy (c.f. Eq.(\ref{eq:TdaggerT})) is based on an expansion in terms of thermal factors.  As mentioned earlier, we need to use resummed propagators in order to deal with pinch singularities correctly. Because of this fact, Eq.(\ref{eq:TdaggerT}) is not useful in a practical sense.  It must be revisited using resummed perturbation theory.  Moreover, to reproduce ``Boltzmann'' Wightman's functions (c.f. Eq.(\ref{eq:PiBoltz})), it is convenient to modify the original expansion so that thermally weighted initial and final states appear more naturally. Thus, we follow a reasoning analogous to the one that led to Eq.(\ref{eq:TdaggerT}) but using resummed propagators and organizing the perturbation series differently.    In order to do that, we rewrite the propagators $G_{12}$ and $G_{21}$ as
\be
G_{12}(p)
&=&
%\theta(p^0)\,(1+n(|p^0|))\,\rho(p) - \theta(-p^0)\,n(|p^0|)\,\rho(p)
\theta(p^0)\,[1+n(|p^0|)]\,\rho_+(p) +
\theta(-p^0)\,n(|p^0|)\,\rho_+(p)
\label{eq:G12r}
\\
G_{21}(p)
&=&
%-\theta(-p^0)\,(1+n(|p^0|))\,\rho(p) + \theta(p^0)\,n(|p^0|)\,\rho(p)
\theta(-p^0)\,[1+n(|p^0|)]\,\rho_+(p) +
\theta(p^0)\,n(|p^0|)\,\rho_+(p)
%= G_{12}(-p)
\label{eq:G21r}
\ee
with
\be
\rho_+(p) &=& {\rm sgn}(p^0)\,\rho(p) = \rho(|p^0|, \bfp)
\ee
\begin{figure}[t]
\begin{center}
\includegraphics[width=0.9\tw]{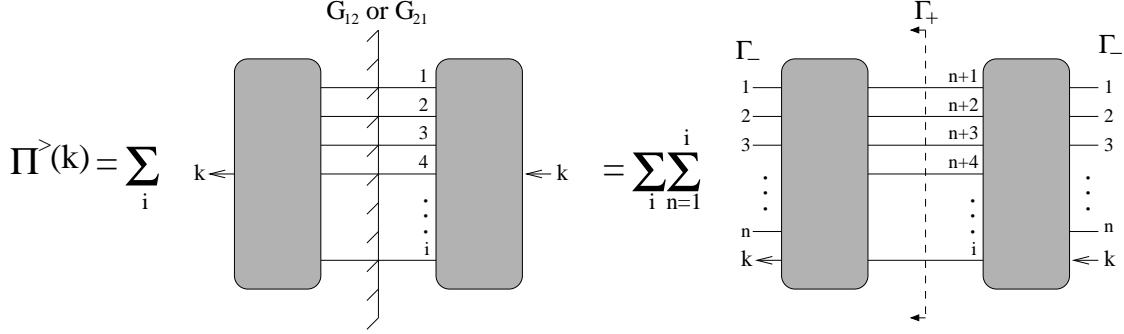}
\end{center}
\caption{The first diagram represents the sum over all skeleton diagrams that have $i$ cut lines. The blob on the shaded side is made of $G_{22}(p)$'s while the blob on the unshaded side is made of $G_{11}(p)$'s. Cut lines are associated with the cut propagators $G_{12}(p)$ and $G_{21}(p)$. The second diagram is obtained from the first by expanding the cut propagators using Eq. (\ref{Eq:phase_vs_prop}) and by changing the variable so that $\theta (-p^{0}) \rightarrow \theta (p^{0})$. Thus, all the states weighted by $d\Gamma^{+}$ becomes final states while the states weighted by $d\Gamma^{-}$ becomes initial states. The lines cut by the dashed line are final states. The arrows indicate the complex conjugated side. We neglected symmetry factors for simplicity.}
\label{fig:Mul_ex}
\end{figure}
The $[1+n(|p^0|)]\rho_+(p)$ factor in $G_{12}$ and $G_{21}$ is naturally associated with a {\em final} state momentum in any given scattering process, and the $n(|p^0|)\rho_+(p)$ factor in $G_{12}$ and $G_{21}$ is associated with an {\em initial} state momentum in any given scattering process.  The idea is to expand the full self-energy using the propagators (\ref{eq:G12r})-(\ref{eq:G21r}); see Fig. \ref{fig:Mul_ex} for an illustration of this procedure.  Note that since we use resummed propagators, only skeleton diagrams appear in the self-energy.  As usual, skeleton diagrams are defined as diagrams without any self-energy insertions.  The result of this procedure can be written in a very suggestive form:
\begin{eqnarray}
\Pi^{>}(k) & = &
\sum_{n=0}^{\infty}
\sum_{s=0}^{\infty} \frac{1}{n!s!}
\left(\int\prod_{i=1}^{n} d\Gamma_-(l_{i})\right)
\left(\int\prod_{j=1}^{s} d\Gamma_+(p_{j})\right)
\nonumber \\ &  & {} \qquad\qquad\qquad
{} \times
\left.
\langle k,\{l_{i}\}_n|{\cal A}_{k}^\dagger|\{p_{j}\}_s\rangle
\langle \{p_{j}\}_s|{\cal A}_{k}|k,\{l_{i}\}_n\rangle
\right|_{\hbox{\scriptsize skeleton}}
\label{eq:PiG}
\end{eqnarray}
where the resummed phase space factors are given by
\begin{eqnarray}
d\Gamma_-(p) & = &
\frac{d^{4}p}{(2\pi)^{4}}\theta(p^{0})\rho_+(p)n(|p^0|)
\\
d\Gamma_+(p) & = &
\frac{d^{4}p}{(2\pi)^{4}}\theta(p^{0})\rho_+(p)[1+n(|p^0|)]
\end{eqnarray}
with the relation (true for rotationally invariant systems)
\begin{eqnarray}
\label{Eq:phase_vs_prop}
d\Gamma_+(p) + d\Gamma_-(-p) & = & \frac{d^{4}p}{(2\pi)^{4}} G_{12}(p)
\end{eqnarray}
Here $\langle \{p_{j}\}_s|{\cal A}_{k}|k,\{l_{i}\}_n\rangle$
is purely made of $G_{11}$ and
represents the sum of all amputated skeleton diagrams\footnote{Note
that Eq.(\ref{eq:PiG}) contains no spectator, but still includes
clusters.  We deal with the latter in Sect. \ref{sec:Approximations}.}
between a set of initial momenta $\{l_i\}_n$
and a set of final momenta $\{p_j\}_s$.
The factor $\langle k,\{l_{i}\}_n|{\cal A}_{k}^\dagger|\{p_{j}\}_s\rangle$ has the
same interpretation and is purely made of $G_{22}$; however, it must be emphasized that
\be
\calK_A(k,\{l_i\}_n,\{p_j\}_s)
=
\left.
\langle k,\{l_{i}\}_n|{\cal A}_{k}^\dagger|\{p_{j}\}_s\rangle
\langle \{p_{j}\}_s|{\cal A}_{k}|k,\{l_{i}\}_n\rangle
\right|_{\hbox{\scriptsize skeleton}}
\label{eq:calK}
\ee
is not the fully connected matrix element squared that
appears in the Boltzmann collision term. This will be examined 
further in the next section.

%%%%%%%%%%%%%%%%%%%%%%%%%%%%%%%%%%%%%%%%%%%%%%%%%%%%%%%%%%%%%%%%%%%%%%%
\subsection{Approximations}
\label{sec:Approximations}
%%%%%%%%%%%%%%%%%%%%%%%%%%%%%%%%%%%%%%%%%%%%%%%%%%%%%%%%%%%%%%%%%%%%%%%

We know that in order to interpret the right hand side of the
Kadanoff-Baym equation  (c.f. Eq.(\ref{Collision_term})) as the
collision integral in the Boltzmann equation, the self-energy $\Pi^>$
(which appears in the loss term) should have the form shown in
Eq.(\ref{eq:PiBoltz}).  Comparing Eq.(\ref{eq:PiBoltz}) with
Eq.(\ref{eq:PiG}) one can see the following differences:
\bitem
\item[(i)] The momentum integrals in Eq.(\ref{eq:PiBoltz}) are over
on-shell 3-momenta whereas the momentum integrals in
Eq.(\ref{eq:PiG}) are over 4-momenta.

\item[(ii)] There are some \textit{extra} terms in Eq.(\ref{eq:calK}). First,
$\calK_A$ is made of the thermal propagators $G_{11}(q)$ and $G_{22}(q)$ while
$\bra{\{p_j\}} {\cal M} \ket{k, \{l_i\}}$ contains only the vacuum
ones ($G_{F}^{0}(q)$ and $G_{F}^{0*}(q)$). Also, $\calK_A$ is made of
connected diagrams that contains clusters (but no spectators), while
$\bra{\{p_j\}} {\cal M} \ket{k, \{l_i\}}$ contains only the fully
connected
diagrams (see Fig. \ref{fig:T3} for the difference between 
connected and fully connected).

\item[(iii)] There are some \textit{missing} terms in Eq.(\ref{eq:calK}).
The terms of the form $G_{F}^{*}(q) G_{F}(q)$ (see Fig. \ref{fig:A4}e and example in Appendix \ref{app:2to2})
are not part of the skeleton expansion in Eq.(\ref{eq:PiG}), by definition.
These terms were present in the original multiple scattering expansion Eq.(\ref{eq:TdaggerT}), 
so they appear to be lost in the skeleton expansion of resummed propagators. They are
actually included in the resummed phase space factor.
\eitem

Since the Boltzmann equation is only a classical approximation to the
full
Kadanoff-Baym equation, one should not expect that $\Pi_{\rm
Boltz.}^>(k)$
in Eq.(\ref{eq:PiBoltz}) is exactly the same as $\Pi^>(k)$ in
Eq.(\ref{eq:PiG}).
Rather, the question to ask is, ``In what sense is
Eq.(\ref{eq:PiBoltz})
contained in Eq.(\ref{eq:PiG})?''
Physically, the Boltzmann equation should be a good approximation in
the
weakly-interacting dilute-gas limit.  Hence, it should be no surprise
that
the answer to the above question is provided by a quasi-particle
ansatz.  

\begin{figure}[t]
\begin{center}
\includegraphics[width=0.9\tw]{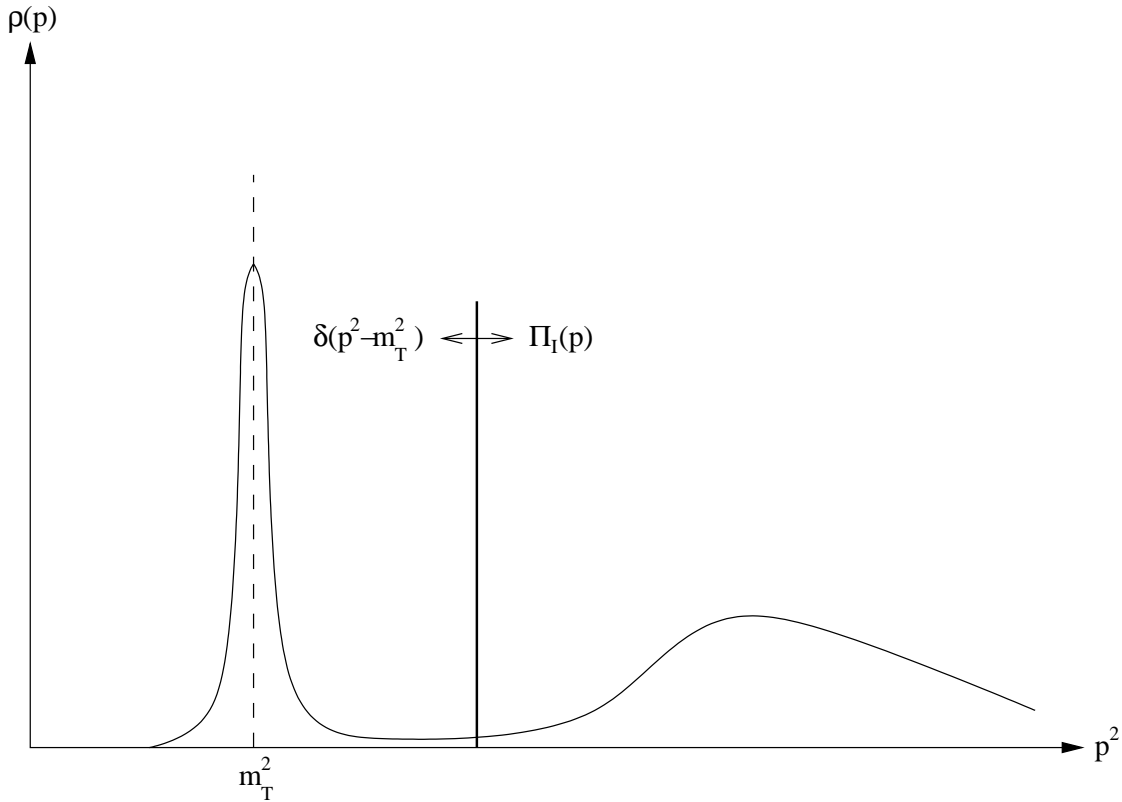}
\end{center}
\caption{
Qualitative depiction of the spectral density. In the quasi-particle
approximation, the left-hand side is approximated by $\delta (p^{2} -
m_{T}^{2})$ while the right-hand side is given by $\Pi_{I}(p)$.
}
\label{fig:spectral_density}
\end{figure}

First consider point (i).
The connection between the 4-momentum and the on-shell 3-momentum is
provided by the following quasi-particle approximation.
In a medium, no momentum state can be absolutely
stable at finite temperature.
However, the Boltzmann equation is an equation for the on-shell
particle
density.  This means that it is valid only when the mean free path is
much
longer than the range of the interaction, which occurs in
weakly interacting systems (i.e. the usual approximations of transport
theory, see Sect. \ref{sec:KB} and \cite{Carrington_etal_2005}).  
This leads to the following condition for the quasi-particle
approximation
\begin{eqnarray}
\label{eq:quasi_particle}
m^{2} - \Pi_{R} \gg |\Pi_{I}^{T}|
\end{eqnarray}
where $\Pi_{I}^{T}$ means the imaginary part of the thermal
self-energy.  If Eq.(\ref{eq:quasi_particle}) holds, the on-shell part
of the spectral
function is well separated from the continuum part (see Fig.
\ref{fig:spectral_density} and \cite{AMY_2003_2,Aarts_Berges_2001}) and
it becomes possible to make the following quasi-particle approximation
\begin{eqnarray}
\label{eq:Spectral_density_quasi}
\rho_+(k) & = & {2\, \Pi_I(|k_0|,k)}\,G_{F}(k) G_{F}^*(k)
\nonumber \\
\rightarrow \rho_{+}^{\rm quasi}(k) & = & 2\pi \delta(k^2 - m_T^2) + {2
\Pi_I(k)}\,G_{V}(k) G_{V}^*(k) \theta(k^2\ne m_T^2)
\end{eqnarray}
where
the $\theta$-function ensures that we do not double-count the on-shell
contribution.
Here $m_{T}^{2} = m^{2} - \Pi_{R}$ is the thermal mass and
we defined
\be
G_{V}(p) = {i\over p^2 - m_T^2 + i\Pi_I^{V}(p)}
\label{eq:GV}
\ee
with
$\Pi_I^{V}$ denoting the imaginary part of the {\em vacuum}
self-energy.  Equations (\ref{eq:Spectral_density_quasi})-(\ref{eq:GV})
constitute our quasi-particle approximation.

Point (ii) is about clusters in $\calK_A$.  The diagrams that contain clusters represent interference between the processes that are occuring within the medium and the scattering process in which the original external momentum $k$ is involved.  For instance, see the diagram on the left hand side of Fig.~\ref{fig:not_include2}.  As such, it is not possible to interpret these terms as a matrix element squared because terms representing independent scattering processes (as shown on the right hand side of Fig.~\ref{fig:not_include2}) are missing.  It is conceivable that one may introduce these missing terms via a multiplicative constant analogous to an effective potential.  But in this work, we need to ignore these interference terms and note that this is a defect of the Boltzmann equation approach.  We simply replace $\calK_A$ with
\be
\calK_A(k,\{l_i\}_n,\{p_j\}_s)
\rightarrow
\left.
\langle k, \{l_{i}\}_n|{\cal M}^\dagger|\{p_{j}\}_s\rangle
\langle \{p_{j}\}_s|{\cal M}|k, \{l_{i}\}_n\rangle
\right|_{\hbox{\scriptsize skeleton}}
\label{eq:replace}
\ee
where $\langle \{p_{j}\}_s|{\cal M}|k, \{l_{i}\}_n\rangle$ now consists solely of fully connected networks of $G_{V}$'s (c.f.~Eq.(\ref{eq:GV})).  Note that the restriction that no self-energy insertion appear still applies.  Note also that ignoring these independent scattering processes (equivalent to neglecting the thermal factors $\Gamma(p)$ in Eqs.(\ref{eq:G11_full})-(\ref{eq:G22_full})) would break gauge invariance in a gauge theory.  As is well-known from the Ward identities for Hard Thermal Loops (HTLs) \cite{Braaten_Pisarski_1990,Frenkel_Taylor_1990}, both propagators and vertices must be HTL resummed to preserve gauge invariance.  In our case, throwing away these independent scattering processes would prevent getting the necessary thermal corrections to the vertices needed for preserving gauge invariance.  We come back to this point and possible solutions in Sect.~\ref{sec:Conclusion}.

Point (iii) is concerned with the problem of generating all the missing
diagrams in our skeleton expansion so that we recover 
full matrix elements squared.  This is a non-trivial problem and we devote
the next section to its solution. In order to make progress, we
need to use the following relationships:
\be
\theta(p^{0}) n(|p^{0}|) 2\Pi_I(p) = \theta(p^{0}) \Pi^<(p)
\label{eq:eq_rels1}
\\
\theta(p^{0})  [1+n(|p^{0}|)]2\Pi_I(p) = \theta(p^{0}) \Pi^>(p)
\label{eq:eq_rels2}
\ee
Equations (\ref{eq:eq_rels1})-(\ref{eq:eq_rels2}) are a rewriting of the usual KMS condition $n(p^{0})\Pi^{>}(p) = (1+n(p^{0}))\Pi^{<}(p)$.  This relation is only valid in equilibrium and makes the collision term of the Kadanoff-Baym equation vanish identically.  However, in an nonequilibrium system, the KMS condition is not satisfied and should not be used.  We argue here that it is still consistent to use Eqs.(\ref{eq:eq_rels1})-(\ref{eq:eq_rels2}) even if we are in a nonequilibrium setting as long as we only use it in Eq.(\ref{eq:iter}) for the computation of $\Pi^{>}(p)$ and $\Pi^{<}(p)$.  The key observation is that $\Pi^{>}(p)$ and $\Pi^{<}(p)$ are proportional to each other {\it up to gradient terms}: this is the essence of the Kadanoff-Baym equation (\ref{Baym_Kadanoff}).  The effect of these gradient terms would be to add higher order gradient terms in our computation of $\Pi^{>}(p)$ and $\Pi^{<}(p)$.  These higher order terms can be neglected, since the Kadanoff-Baym equation is obtained from an expansion to lowest order in the gradients.  In this sense, Eqs.(\ref{eq:eq_rels1})-(\ref{eq:eq_rels2}) are approximate expressions that are consistent with the the Kadanoff-Baym equation.

%%%%%%%%%%%%%%%%%%%%%%%%%%%%%%%%%%%%%%%%%%%%%%%%%%%%%%%%%%%%%%%%%%%%%%%
\section{Collision Term Derivation}
\label{sec:derivation}
%%%%%%%%%%%%%%%%%%%%%%%%%%%%%%%%%%%%%%%%%%%%%%%%%%%%%%%%%%%%%%%%%%%%%%%

As explained in Sect. \ref{sec:Approximations}, we need to show how the
missing terms in the squared scattering matrix elements are generated. 
With the approximations
presented in the last section, Eq.(\ref{eq:PiG}) becomes
\be
\Pi^>(k) & = &
\sum_{n=0}^\infty \sum_{s=0}^\infty \frac{1}{s! \, n!}
\int\prod_{k=1}^{n} \frac{d^4l_{k}}{(2\pi)^{4}}\theta(l_k^0)
\left[n(|l_{k}^{0}|)2\pi\delta(l_{k}^{2} - m_T^2) +
G_{V}^*(l_{k})\Pi^{<}(l_{k})G_{V}(l_{k})\theta(k^2\ne
m_T^2)\right]
\nonumber \\
         &   & \times
\int\prod_{j=1}^{s}
\frac{d^4p_{j}}{(2\pi)^{4}}
\theta(p_j^0)
\left[[1+n(|p_j^0|)]
2\pi\delta(p_j^2 - m_T^2) +
G_{V}(p_j)\Pi^{>}(p_j)G_{V}^*(p_j)\theta(k^2\ne m_T^2)
\right] \nonumber \\
         &   & \times
\left.
\bra{k, \{l_k\}_n}{{\calM}}^\dagger\ket{\{p_j\}_s}
\bra{\{p_j\}_s}{{\calM}} \ket{k, \{l_k\}_n}
\right|_{\rm skeleton}
\label{eq:iter}
\ee
with a similar expression for $\Pi^<(k)$.  This equation defines a
self-consistent equation for $\Pi^>$ and $\Pi^<$.  Our problem is now
reduced to showing that the Boltzmann ansatz, Eq.(\ref{eq:PiBoltz}),
is a solution of Eq.(\ref{eq:iter}).

\begin{figure}[t]
\begin{center}
\includegraphics[width=0.9\tw]{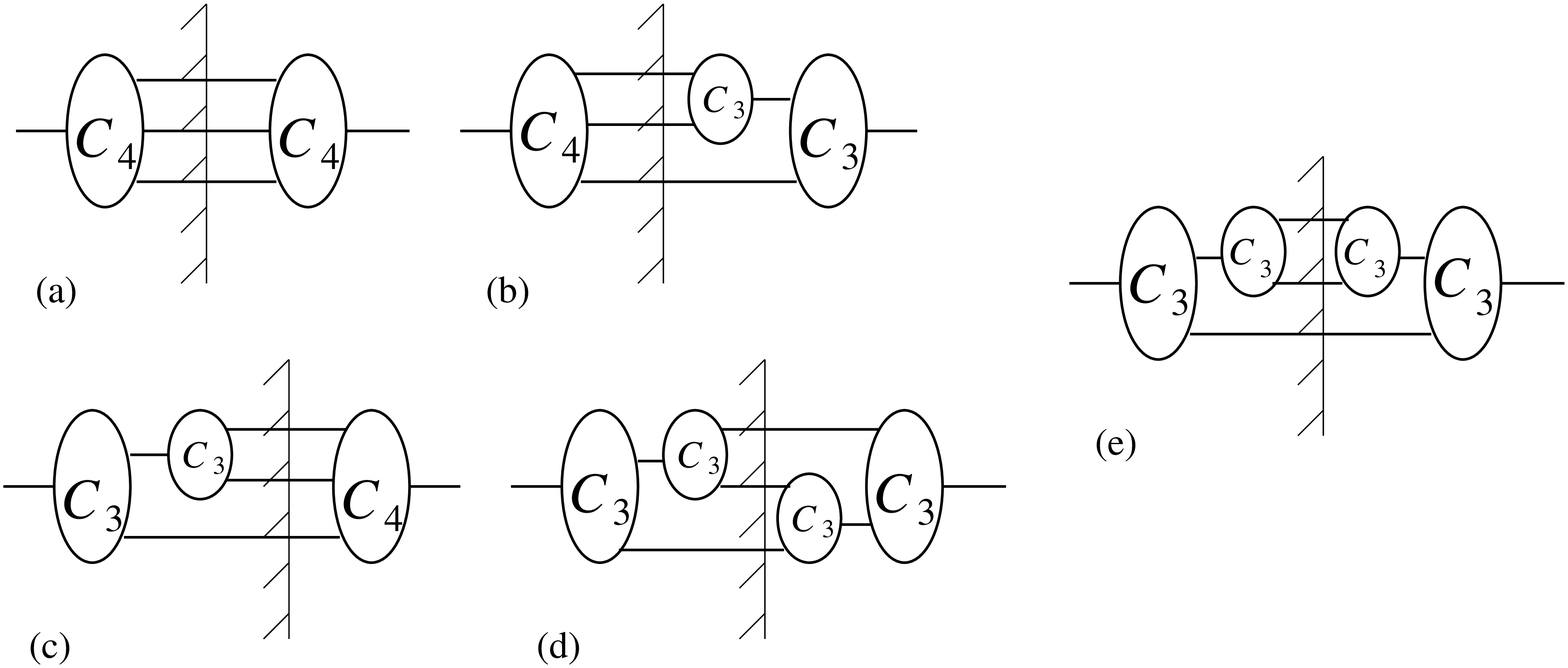}
\end{center}
\caption{
Diagrammatic depiction of $|M_4|^2$.  Here $C_n$ denotes the sum
of all 1-particle irreducible skeleton diagrams with
$n$ external legs.
Diagrams (a-d) make up
$\left.
\bra{k,\{l_k\}}{{\calM}}^\dagger\ket{\{p_j\}}
\bra{\{p_j\}}{{\calM}} \ket{k,\{l_k\}}
\right|_{\rm skeleton}$ while
(e) is generated by lower order (in the number of cut momenta)
skeleton diagram as shown in Fig.~\ref{fig:P4}.
}
\label{fig:A4}
\end{figure}

Qualitatively, we can argue as follows:
the scattering matrix element
$\bra{\{p_j \}}{{\calM}}\ket{k,\{l_k\}}$
is now made of zero temperature Feynman
propagators $G_{V}(p)$.
However, we cannot yet say that
\be
\calK_M
=
\left.
\bra{k, \{l_k\}_n}{{\calM}}^\dagger\ket{\{p_j\}_s}
\bra{\{p_j\}_s}{{\calM}} \ket{k, \{l_k\}_n}
\right|_{\rm skeleton}
\ee
equals 
$
\left|\bra{\{p_j\}_s}{{\calM}}\ket{k, \{l_k\}_n}\right|^2
$.
This is again because
diagrams in the skeleton expansion
cannot include self-energy insertions and
hence do not contain all Feynman diagrams contributing to
$\left|\bra{\{p_j\}_s}{{\calM}}\ket{k, \{l_k\}_n}\right|^2$.
We should emphasize here that the matrix element
$\bra{\{p_j\}_s}{{\calM}}\ket{k,\{l_k\}_n}$
itself {\em does} contain
all necessary diagrams. It is the joining of the final state momenta
$\{p_j\}_s$ that makes the distinction between
$\left|\bra{\{p_j\}_s}{{\calM}}\ket{k, \{l_k\}_n}\right|^2$ and
$\calK_M$.
This is illustrated in Fig.~\ref{fig:A4} for the part
of $\Pi^>$ which involves 4 particle (2-2 or 1-3) processes.
The same figure also illustrates that the missing diagrams are the ones
with
self-energy insertions.

\begin{figure}[t]
\begin{center}
\includegraphics[width=0.8\tw]{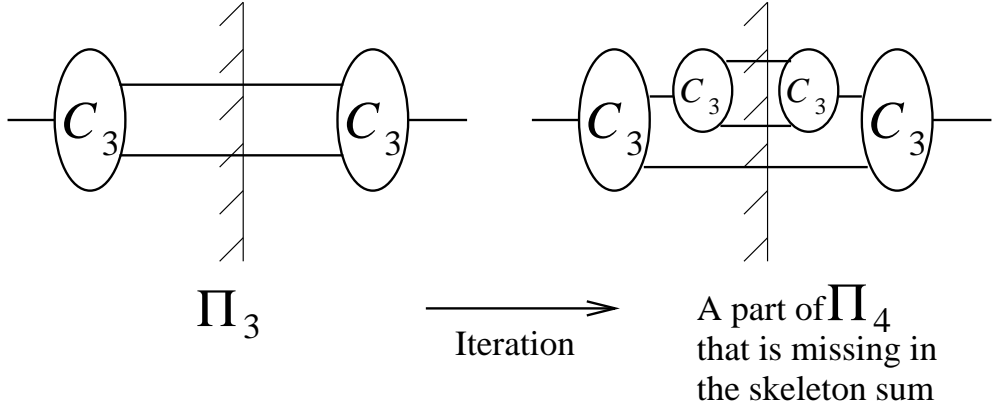}
\end{center}
\caption{
Lower order (in the number of the cut momenta) skeleton diagram that
generate the 2-particle reducible diagram (e) in Fig.~\ref{fig:A4} when
iterated once.
}
\label{fig:P4}
\end{figure}

\begin{figure}[t]
\begin{center}
\includegraphics[width=0.7\tw]{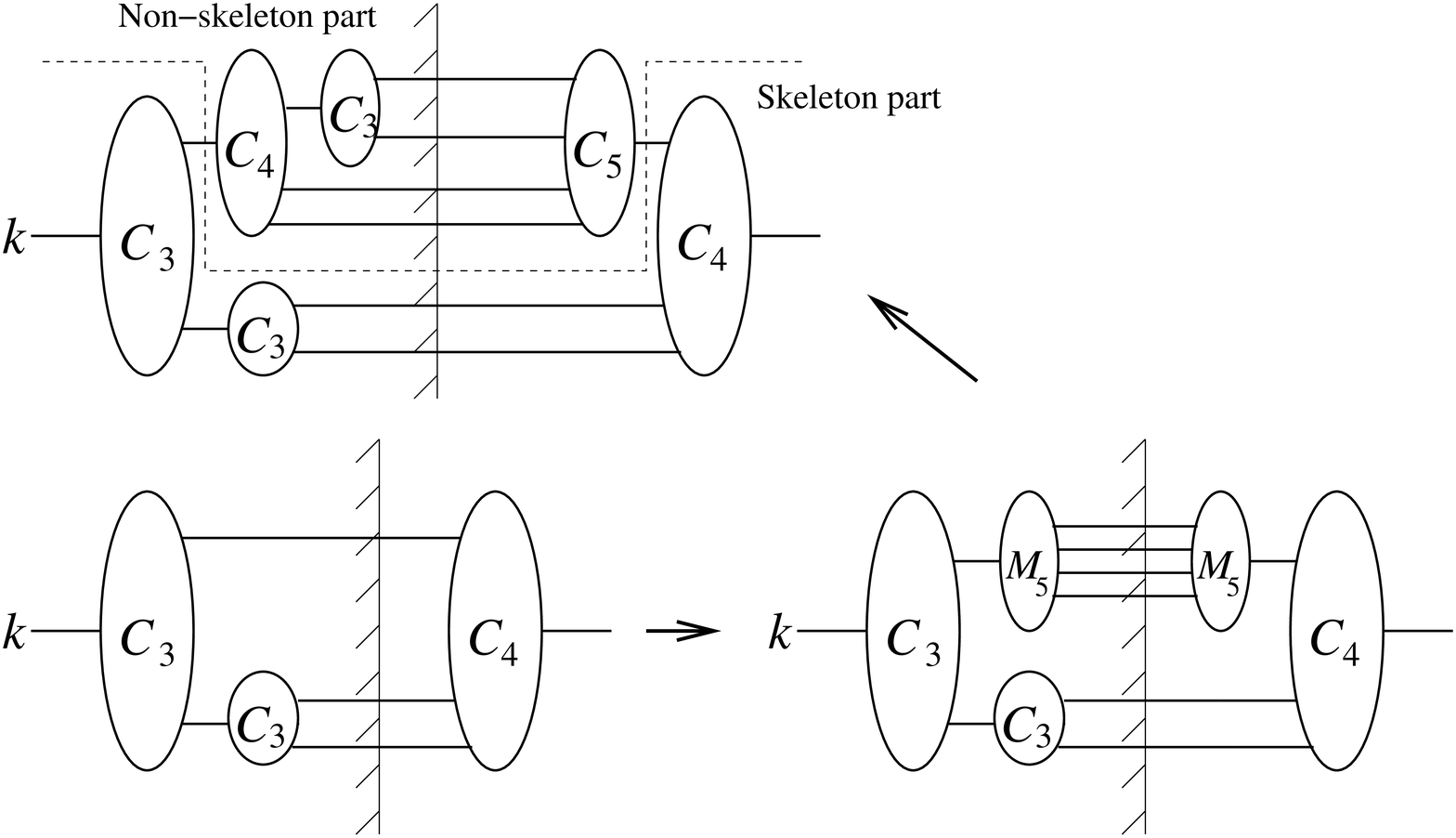}
\end{center}
\caption{A nested non-skeleton diagram that contributes to $|M_7|^2$.
Also shown are the lower order skeleton diagram which generates the
nested
non-skeleton diagram when iterated with $M_5$.
}
\label{fig:nested}
\end{figure}

Since the right hand side of Eq.(\ref{eq:iter})
includes
$\Pi^{>}$ and $\Pi^{<}$, this equation must be solved iteratively; hence, the missing diagrams must come from this iteration procedure.
In other words, diagrams missing in the skeleton sum
$\calK_M$ should all be generated by the lower order (in the number
of legs) skeleton diagrams by trading some on-shell
$\delta$-functions with $G_{V}^* \Pi^{>,<} G_{V}$ combinations.
For example, the diagram labelled (e) in Fig.~\ref{fig:A4} is generated
by
one iteration as shown in Fig.~\ref{fig:P4}.
Fig.~\ref{fig:nested} contains a more elaborate example.
Hence, one can at least see that Eq.(\ref{eq:PiBoltz}) and
Eq.(\ref{eq:iter}) contain the same set of diagrams.  If the symmetry
factors all work out, then we have shown that the two expressions are
equivalent.  An example of this procedure involving
2 to 2 scattering processes is presented in Appendix~\ref{app:2to2}.

\begin{figure}[t]
\begin{center}
\includegraphics[width=0.9\tw]{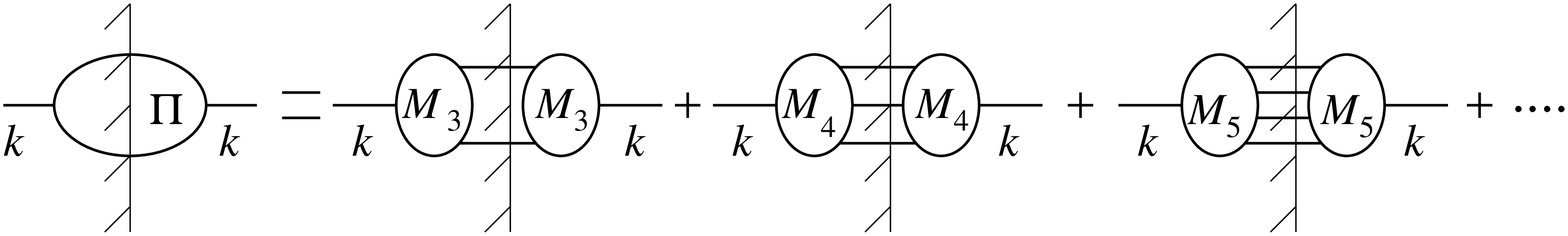}
\end{center}
\caption{
Graphical representation of Eq.(\ref{eq:PiBoltz2}).
}
\label{fig:self_expansion}
\end{figure}

Formally, what we need to show is that Eq.(\ref{eq:PiBoltz}) is the
iterated solution of Eq.(\ref{eq:iter}).
To do so, we define
\be
d\Gamma_-^0(p) & = &
\frac{d^{4}p}{(2\pi)^{4}}\theta(p^{0})n(|p^0|)2\pi\delta(p^2 - m_T^2)
\\
d\Gamma_+^0(p) & = &
\frac{d^{4}p}{(2\pi)^{4}}\theta(p^{0})
[1+n(|p^0|)]2\pi\delta(p^2-m_T^2)
\ee
and
\be
\label{eq:phase_space}
d\Gamma^0(p) =
d\Gamma_+^0(p) + d\Gamma_-^0(-p)
=
\frac{d^{4}p}{(2\pi)^{4}} G_{12}^0(p)
\ee
and rewrite Eq.(\ref{eq:PiBoltz}) as follows
\be
\Pi_{\rm Boltz}^>(k)
& = &
\sum_{n=2}^\infty
{1\over n!}
\left(\int \prod_{i=1}^n
\,d\Gamma^0(l_{i})\right)\,|M_{n+1}(k,\{l_i\}_n)|^2
\label{eq:PiBoltz2}
\ee
where $M_{n+1}(k,\{l_i\}_n)$
is the sum of all Feynman diagrams involving $k$ and a set of
$n$ external momenta $\{l_i\}_n$.  Depending on the sign of the energy,
an $l_i$ can be either an initial-state momentum or a final-state
momentum.
A graphical representation of Eq.(\ref{eq:PiBoltz2}) is given in
Fig.~\ref{fig:self_expansion}.  Note here that a similar result has been obtained in the two-loop case \cite{Majumder_Gale_2002}.
Using the symmetry of the bosonic multi-particle state,
we also rewrite Eq.(\ref{eq:iter}) as
%%%%%%%%%%%%%%%%%%%%%%%%%%%%%%%%%%%%%%%%%%%%
%
\be
\Pi^>(k) & = &
\sum_{n=0}^\infty \sum_{s=0}^\infty
{1\over n!\,s!}
\non
& & {} \times
\sum_{m=0}^n
{n!\over m! (n-m)!}
\left(\int
\prod_{a=1}^{n-m} d\Gamma_-^0(l_a)\right)
\left(\int
\prod_{b=1}^{m}
{d^4 \tilde{l}_b\over (2\pi)^4}\,\theta(l_b^0)\,G_{V}(\tilde{l}_b)G_{V}^*(\tilde{l}_b)
2\Pi^<(\tilde{l}_b)\right)
\non & & {} \times
\sum_{t=0}^s
{s!\over t! (s-t)!}
\left(\int
\prod_{c=1}^{s-t} d\Gamma_+^0(p_c)\right)
\left(\int
\prod_{d=1}^{t}
{d^4 \tilde{p}_d\over (2\pi)^4}\,\theta(p_d^0)\,G_{V}(\tilde{p}_d)G_{V}^*(\tilde{p}_d)
2\Pi^>(\tilde{p}_d)\right)
\non & & {} \times
\left.
\bra{k, \{l_a, \tilde{l}_b\}_n}{{\calM}}^\dagger\ket{\{p_c,
\tilde{p}_d\}_s}
\bra{\{p_c, \tilde{p}_d\}_s}{{\calM}} \ket{k, \{l_a, \tilde{l}_b\}_n}
\right|_{\rm skeleton}
\label{eq:iter2}
\ee
where the tilde over a momentum variable is a reminder that it is an
off-shell momenta.  Substituting Eq.(\ref{eq:PiBoltz2}) into the right
hand side of Eq.(\ref{eq:iter2}) and
shifting some indices, one arrives at
\be
\lefteqn{\Pi^>(k)\, =\,
\sum_{n=0}^\infty
\sum_{m=0}^\infty
\sum_{t=0}^\infty
\sum_{s=0}^\infty
}
&&
\non
& & {} \times
{1\over m! n!}
\int
\prod_{a=1}^{n} d\Gamma_-^0(l_a)
\int
\prod_{b=1}^{m}
{d^4 \tilde{l}_b\over (2\pi)^4}\,\theta(l_b^0)\,G_{V}(\tilde{l}_b)G_{V}^*(\tilde{l}_b)
\non & & {} \qquad\times
\left[
\sum_{u_b=2}^\infty
{1\over u_b !}
\int \prod_{i_{u_b}=1}^{u_b} d\Gamma^0(l_{i_{u_b}})\,
|M_{u_b+1}(\tilde{l}_b,\{l_{i_{u_b}}\}_{u_b})|^2
\right]
%\non & & {} \qquad\times
%\left[
%\sum_{\sigma_b=0}^\infty \sum_{\rho_b=0}^\infty
%{1\over \sigma_b!\, \rho_b!}\,
%\int \prod_{u_b=1}^{\sigma_b} d\Gamma_-^0(l_{u_b})\,
%\int \prod_{v_b=1}^{\rho_b} d\Gamma_+^0(p_{v_b})\,
%\left|
%\bra{l_b, \{p_{v_b}\}_{\rho_b}}\calM \ket{\{l_{u_b}\}_{\sigma_b}}
%\right|^2
%\right]
%%
\non & & {} \times
{1\over t! s!}
\int
\prod_{c=1}^{s} d\Gamma_+^0(p_c)
\int
\prod_{d=1}^{t}
{d^4 \tilde{p}_d\over (2\pi)^4}\,\theta(p_d^0)\,G_{V}(\tilde{p}_d)G_{V}^*(\tilde{p}_d)
\non & & {} \qquad \times
\left[
\sum_{v_d=2}^\infty
{1\over v_d !}
\int \prod_{j_{v_d}=1}^{v_d} d\Gamma^0(l_{j_{v_d}})\,
|M_{v_d+1}(\tilde{p}_d,\{l_{j_{v_d}}\}_{v_d})|^2
\right]
%\non & & {} \qquad \times
%\left[
%\sum_{\alpha_d=0}^\infty \sum_{\beta_d=0}^\infty
%{1\over \alpha_d!\, \beta_d!}\,
%\int \prod_{i_d=1}^{\alpha_d} d\Gamma_-^0(l_{i_d})\,
%\int \prod_{j_d=1}^{\beta_d} d\Gamma_+^0(p_{j_d})\,
%\left|
%\bra{\{p_{j_d}\}_{\beta_d}}\calM \ket{p_d, \{l_{i_d}\}_{\alpha_d}}
%\right|^2
%\right]
\non & & {} \times
\left.
\bra{k, \{l_a, \tilde{l}_b\}_{n+m}}{{\calM}}^\dagger\ket{\{p_c,
\tilde{p}_d\}_{s+t}}
\bra{\{p_c, \tilde{p}_d\}_{s+t}}{{\calM}} \ket{k,\{l_a,
\tilde{l}_b\}_{n+m}}
\right|_{\rm skeleton}
\label{eq:solution}
\ee
What we need to show is that
Eq.(\ref{eq:PiBoltz2}) and Eq.(\ref{eq:solution}) are in fact
equivalent.

\begin{figure}[t]
\begin{center}
\includegraphics[width=0.8\tw]{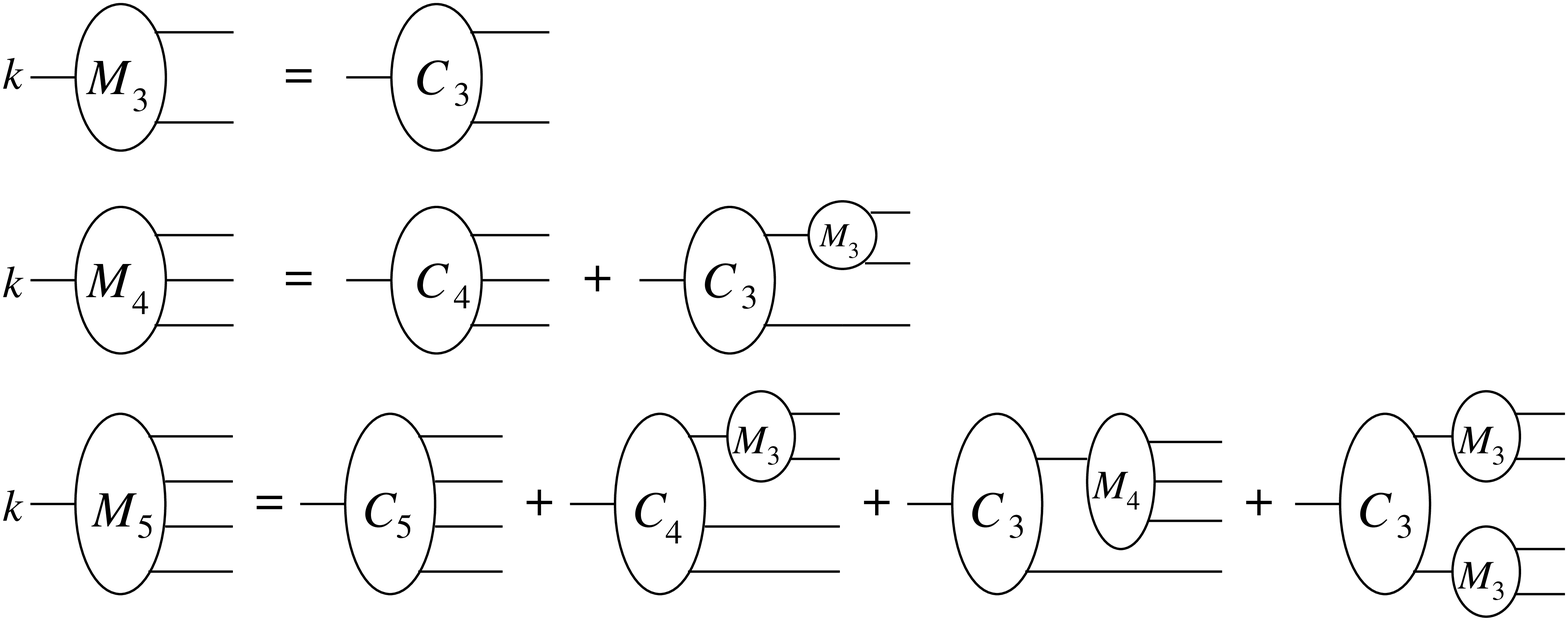}
\end{center}
\caption{
Decomposition of $M_n$ in
terms of the 1-particle irreducible part $C_n$
and other $C_m$ and $M_m$ with $m < n$.
Here $n$ and $m$ indicate
the number of external momenta involved in the process including
$k$.
For the process involving 3 external momenta, we have $M_3 = C_3$.
The $M_4$ that appears in the third line can be further expanded using
the
second line.
}
\label{fig:M4}
\end{figure}

To show that the two expressions are indeed equivalent,
we start with Eq.(\ref{eq:PiBoltz2}).
Consider an $M_{n}$ in Eq.(\ref{eq:PiBoltz2}).
What we would like to do now is to
decompose $M_{n}$ in terms of 1-particle irreducible parts.
Since Eq.(\ref{eq:PiBoltz2}) is the result of the skeleton expansion,
the lowest order (in the number of legs) $M_{n}$ is $M_3$ and has only the 1-particle irreducible part shown in
Fig.~\ref{fig:M4}.
For $n > 3$, $M_n$ can be
decomposed into the 1-particle irreducible diagram
involving all $n$ particles (denoted by $C_n$)
and diagrams that are combinations of lower order
$C_m$'s and $M_{m'}$'s with $m < n$ and $m' < n$ as illustrated in
Fig.~\ref{fig:M4}.
The $M_{m'}$ part in turn is
made of lower order $C$'s and $M$'s, and the iteration continues
until no $M$ is left.

Since the iteration starts
with $M_3 = C_3$, one can then say that $M_n$ consists of all tree-like
diagrams but with the vertices replaced by appropriate $C_m$ with $m
\le n$.
These diagrams, however, are not the tree diagrams of the underlying
theory since the tree diagrams in the underlying scalar field
theory can only have
3-momentum and 4-momentum vertices while in these tree-like diagrams
any
$C_m$ can play the role of a vertex.
As an example, we show graphical representation of this decomposition
for
$M_3, M_4$ and $M_5$ in Fig.~\ref{fig:M4}.

The diagrams that contribute to $|M_n|^2$ can be now generated
by summing over all pairs of the tree-like diagrams from $M_n$ and
$M_n^*$
and over all possible distinct joinings of the $n-1$ legs
(excluding $k$) between the chosen pair; see Fig.~\ref{fig:A4} for an example.
Among these joined diagrams, the majority are skeleton diagrams.
A non-skeleton diagram results when
a tree branch from the $M_n$
side fully connects with a tree branch from the $M_n^*$ side as
in the diagram (e) of Fig.~\ref{fig:A4}.
These branches could be a primary branch which connects to
the root $C$
(where the external momentum $k$ enters) by a
single line,
or an $r$-th removed branch which needs to go through at least
$r$ number of $C$'s to join with the root $C$.
Since all possible tree-like diagrams
are being summed over,
it is clear that
these joined branches sum to $|M_{q+1}|^2$, where $q$ is the number
of joined legs.
It is also clear that  replacing all joined branches
with the on-shell part in
Eq.(\ref{eq:Spectral_density_quasi}) reduces the diagram to a lower
order skeleton diagram and this reduction is unique.
Consequently,
\be
\lefteqn{|M_{N+1}(k, \{l_i\}_N)|^2 \, =
\sum_{m=0}^\infty
\sum_{n=0}^\infty
\sum_{t=0}^\infty
\sum_{s=0}^\infty
\,}
&&
\non
& & {}
{1\over m! n!}
\left(\int
\prod_{b=1}^{m}
{d^4 \tildel_b\over (2\pi)^4}\,\theta(l_b^0)\,G_{V}(\tildel_b)
G_{V}^*(\tildel_b)\right)
%\non & & {} \qquad\times
\left[
\sum_{u_b=2}^\infty
{1\over u_b !}
|M_{u_b+1}(\tildel_b,\{l_{i_{u_b}}\}_{u_b})|^2
\right]
\non & & {} \times
{1\over t! s!}
\left(\int
\prod_{d=1}^{t}
{d^4 \tildel_d\over (2\pi)^4}\,\theta(\tildel_d^0)\,G_{V}(\tildel_d)
G_{V}^*(\tildel_d)\right)
%\non & & {} \qquad \times
\left[
\sum_{v_d=2}^\infty
{1\over v_d !}
|M_{v_d+1}(\tildel_d,\{l_{j_{v_d}}\}_{v_d})|^2
\right]
\non & & {} \times
\left.
\bra{k,\{l_a, \tildel_b\}_{n+m}}{{\calM}}^\dagger
\ket{\{l_c, \tildel_d\}_{s+t}}
\bra{\{l_c, \tildel_d\}_{s+t}}{{\calM}} \ket{k,\{l_a, \tildel_b\}_{n+m}}
\right|_{\rm skeleton}
\non & & {} \times
\delta(N - n - s - u_b - v_d)
\label{eq:MN}
\ee
with the external momenta given by
$
\{l_i\}_N
=
\{l_a, l_c, l_{i_{u_b}}, l_{j_{v_d}}\}_{n+s+u_b+v_d}
$ and
$\delta(N - n - s - u_b - v_d)$ is a Kronecker-$\delta$.
The symmetry factors in Eq.(\ref{eq:MN}) are obtained as follows.
One starts with
\be
{1\over (n+m)! (s+t)!}
\left.
\bra{k,\{l_i\}_{n+m}}{{\calM}}^\dagger\ket{\{l_f\}_{s+t}}
\bra{\{l_f\}_{s+t}}{{\calM}} \ket{k,\{l_i\}_{n+m}}
\right|_{\rm skeleton}
\ee
where the factorial factors accounts for the bosonic symmetry.
For the initial momenta $\{l_i\}_{n+m}$,
there are $(n+m)!/(m!n!)$ number of ways to choose $m$ legs that
receives a self-energy insertion.
Combining with the original $1/(n+m)!$, this
yields
the $1/(m!n!)$ factor in the second line.
The $1/(s!t!)$ factor in the third line is similarly obtained.
The $1/u_b!$ and $1/v_d!$ factors come from the fact that  the
symmetry factor for a diagram is given by the order of the
permutation group that leaves the diagram intact
\cite{Jeon_Ellis_1998,Weinstock_2005}.

Eq.(\ref{eq:MN}) is exactly the content of Eq.(\ref{eq:solution}). We
have
thus proved that Eq.(\ref{eq:PiBoltz2}) is the iterative solution of
Eq.(\ref{eq:iter}).

So far, no equilibrium
assumption has entered into our derivation except for
Eqs.(\ref{eq:eq_rels1}) and (\ref{eq:eq_rels2}).
Hence it is possible to replace the
bosonic equilibrium distribution function $n(|p^{0}|)$ by its
non-equilibrium
counterpart $f(X,\mathbf{p})$.
Within our quasi-particle approximation, the Boltzmann collision
term is given by
\be
\lefteqn{C(X, k) \, = \,}
&&
\non
& & {}
{1\over 2}[1 + f(X, \bfk)]
\sum_{s=0}^{\infty} \sum_{n=0}^{\infty} \frac{1}{n!\,s!}
\left(\int\prod_{i=1}^{n}\frac{d^3 l_i}{(2\pi)^3 2E_i} f(X,
\bfl_i)\right)
\left(\int\prod_{j=1}^{s}\frac{d^3 p_j}{(2\pi)^3 2E_j} [1+f(X,
\bfp_j)]\right)
\non & & {} \qquad\qquad\qquad\qquad\qquad\qquad \times
%\non & & {} \qquad\qquad\qquad\qquad\qquad\qquad {}\times
\big| \langle k, \{ p_j \}_s | {\cal M} | \{l_i\}_n \rangle \big|^2
\non
& & {}
-
{1\over 2} f(X, \bfk)
\sum_{s=0}^{\infty} \sum_{n=0}^{\infty} \frac{1}{n!\,s!}
\left(\int\prod_{i=1}^{n}\frac{d^3 l_i}{(2\pi)^3 2E_i} f(X,
\bfl_i)\right)
\left(\int\prod_{j=1}^{s}\frac{d^3 p_j}{(2\pi)^3 2E_j} [1+f(X,
\bfp_j)]\right)
\non & & {} \qquad\qquad\qquad\qquad\qquad\qquad \times
%\non & & {} \qquad\qquad\qquad\qquad\qquad\qquad {}\times
\big| \langle \{ p_j \}_s | {\cal M} |k, \{l_i\}_n \rangle \big|^2
\label{eq:main_result}
\ee
where
$
\big| \langle \{ p_j \} | {\cal M} |k, \{l_i\} \rangle \big|^2
$
is the full scattering matrix squared in vacuum with zero temperature
masses replaced by thermal masses.
The sum $n+s$ should, of course, be greater than or equal to 2 and the
energy-momentum conserving $\delta$-function is absorbed in the
definition
of matrix element squared.  This result is similar to the one in Ref.
\cite{Weinstock_2005}, with the difference that 
in Ref.\cite{Weinstock_2005}
only tree diagrams are included in the calculation of $|M|^2$.
The analysis in Ref.\cite{Weinstock_2005} also did not fully include
resummed propagators.

The main feature of Eq.(\ref{eq:main_result})
is that it is
naturally  expressed in terms of full scattering matrix elements.
They can be evaluated theoretically using perturbation theory or by
using
data from experiments.  Note also that all the information about the
theory is in the matrix elements. Therefore, the expressions
of the gain and loss rates should be valid for any interacting theory 
of a  real scalar field.
In the case of $\phi^3$ and $\phi^4$, it
reproduces the results of \cite{Carrington_etal_2005} for 1,2,3,4-loop
contributions at tree level.
In principle, our analysis could be generalized to gauge theories but 
the preservation of gauge invariance must be carefully dealt with.

%%%%%%%%%%%%%%%%%%%%%%%%%%%%%%%%%%%%%%%%%%%%%%%%%%%%%%%%%%%%%%%%%%%%%%%
\section{Discussions and Conclusions}
\label{sec:Conclusion}
%%%%%%%%%%%%%%%%%%%%%%%%%%%%%%%%%%%%%%%%%%%%%%%%%%%%%%%%%%%%%%%%%%%%%%%

Equations Eq.(\ref{eq:PiBoltz}) and
(\ref{eq:main_result}) are our main results in this paper.
Namely, we have shown that the Wightman self-energy can be
approximately
expressed in terms of the thermally averaged full
scattering matrix elements.
We then used this fact to express the Boltzmann collision term to all
orders.
Equations Eq.(\ref{eq:PiBoltz}) and
(\ref{eq:main_result})
are often used in various calculations such as
in-medium particle property calculations
\cite{Gale:2005ri,Mallik:2001qa,
Eletsky:2001bb,
Kapusta:2001jw, Amelino-Camelia:1999yr, Gao:1998mn, Eletsky:1998jm}
and Monte-Carlo simulations of many-body systems; but, as far as we know, a systematic derivation has been lacking.
Within our quasi-particle approximation, our result is valid for
all orders of perturbation theory and any number of participating
particles.

Our final results are quite independent of the details of the
underlying
theory because they involve only the scattering matrix elements.
Even though our discussions and results are for
real scalar theories, they can in principle
be generalized to more complicated theories such as gauge theories.
This is clearly the subject of a future work.

An advantage of having such an all-orders formulation
is that these matrix
elements do not have to be calculated from first principles; instead, they can
be measured in vacuum and used as an input for the in-medium calculation.
Also, having derived Eq.(\ref{eq:main_result}), we can use it as a basis
for
Monte-Carlo simulations of many-body systems with confidence.

While obtaining our main results, we have made some approximations.
Here we would like to list them and comment on them.
\bitem

\item[(i)] Approximations in deriving the Kadanoff-Baym equation.

These approximations are valid in the weakly inhomogeneous system and 
the weak interaction strength limit. 
Since the Kadanoff-Baym equation has been much
discussed in the literature (see for example \cite{Juchem:2004cs,Lindner:2005kv,Knoll:2001jx,Ivanov:1998nv} for some discussion in the context of 2PI effective action methods), we have little more to add to this point. 

\item[(ii)] Quasi-particle approximation of the spectral density,
    Eq.(\ref{eq:Spectral_density_quasi}).

This is somewhat different than the quasi-particle ansatz made in
\cite{Carrington_etal_2005,Weinstock_2005} in that we have included the
continuum part of the spectral density.  This approximation should be
valid under the same conditions that the Kadanoff-Baym equation is
valid.  To go beyond that, we need to repeat our analysis using the
off-shell transport theory developed in Refs.~\cite{Landsman_VanWeert_1987,
Landsman:1988ta, Leupold:1999ga, Ivanov:1999tj, Juchem:2004cs, Cassing:2000ch}.
This is clearly outside the scope of the present paper.

\item[(iii)] Ignoring thermal corrections to the interaction vertices,
Eq.(\ref{eq:replace}).

Our quasi-particle ansatz could easily accomodate the in-medium mass,
but we had to make an approximation where we ignored thermal vertex
corrections.
 Physically, this amounts to ignoring interference between the purely
 in-medium processes and processes in which the external momentum
 $k$ is involved.
 In many physical situations, ignoring these corrections
 can lead to serious inconsistencies.
 In the case of real scalar theories, this is
 not much of a concern. But when generalizing to gauge theories, thermal vertex
 corrections must be taken into account.
 For instance, in hard thermal loop calculations in hot QCD, it is
 essential to include thermal corrections to both vertices and
 self-energies to preserve gauge symmetry.

 It should be possible generalize our formulation to include
 thermal vertex corrections, but we will have to pay a price: the scattering
matrix
 elements need to include disconnected pieces so that both diagrams in
 Fig.~\ref{fig:not_include2} are included.
 Suppose we denote the sum of all cut in-medium diagrams (the analogue
of
 the vacuum bubbles) by $W_T$.  In vaccum, this is zero because the vacuum
 cannot spontaneously generate propagating particles.  In a medium,
however,
 $W_T$ encodes contributions from
 the independent scattering processes occuring in the medium.
 Just like the vacuum phase, the missing diagrams in $\calK_A$ can be
 added by multiplying the self-energy by $e^{W_T}$.
 But this extra factor is not a part of the self-energy. Hence, we must
 compensate it with an additional factor of $e^{-W_T}$.
 Whether we can repeat our analysis in this way is currently under
 investigation.

\eitem

In summary, we have greatly extended our previous work
\cite{Gagnon_etal_2005} and a similar
work by Weinstock \cite{Weinstock_2005} to derive the Boltzmann
collision term from the Kadanoff-Baym
equation as fully as possible.  Along the way, we have identified the
approximations that have to be made and possible remedies to include
the ignored effects. Work along this line is continuing.
It is hoped that this work lays a firm foundation for the inclusion of
many-body scattering processes in Monte-Carlo simulations.  We also mention that it is possible to generalize this result to include strong classical sources.  First steps in this direction have been taken in Ref. \cite{Gelis_Venugopalan_2006} and work along these lines is currently in progress.

%%%%%%%%%%%%%%%%%%%%
% Acknowledgements %
%%%%%%%%%%%%%%%%%%%%

\begin{acknowledgments}
We thank S. Turbides for useful discussions and A. Berndsen for a careful proofreading of the manuscript.  This work was supported
in part
by
the Natural Sciences and Engineering Research Council of Canada, and in
part,
for S.J.,
by le Fonds Nature et Technologies du Qu\'ebec.  S.J.~also
thanks RIKEN BNL Center and U.S. Department of Energy
[DE-AC02-98CH10886] for
providing facilities essential for the completion of this work.
\end{acknowledgments}

%%%%%%%%%%%%%%%%%%%%%%%%%%%%%%%%%%%%%%%%%%%%%%%%%%%
%%%%%%%%%%%%%%%%%%%%%%%%%%%%%%%%%%%%%%%%%%%%%%%%%%%
%%%%%%%%%%%%%%%%%%%%%%%%%%%%%%%%%%%%%%%%%%%%%%%%%%%

%%%%%%%%%%%%%%%%
% Bibliography %
%%%%%%%%%%%%%%%%

%%%%%%%%%%%%%%%%%%%%%%%%%%%%%%%%%%%%%%%%%
%              APPENDICES
%%%%%%%%%%%%%%%%%%%%%%%%%%%%%%%%%%%%%%%%%
\appendix

\section{2 $\leftrightarrow$ 2 process example}
\label{app:2to2}

\begin{figure}[t]
\begin{center}
\includegraphics[width=0.7\tw]{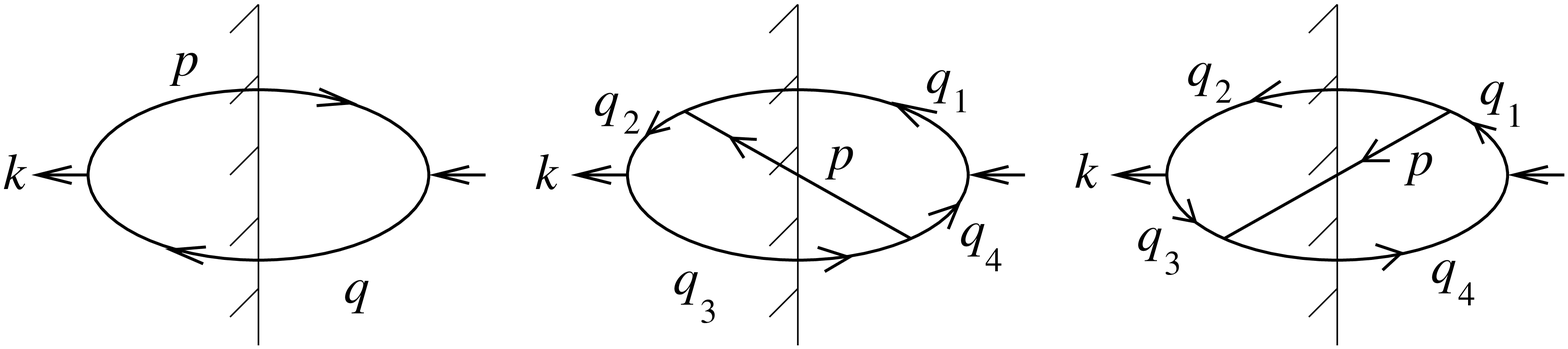}
\end{center}
\caption{
Lowest order skeleton diagrams in the $g\phi^3$ theory.
}
\label{fig:lowest}
\end{figure}

In this example, we illustrate
how the iteration procedure described
in Sect. \ref{sec:derivation} adds necessary diagrams to get full
scattering matrix elements using the $g\phi^3$ scalar theory.
We will look at the first non-trivial and
kinematically allowed process: the 2$\rightarrow$2 process. We start
from the skeleton expansion of the Wightman self-energy.
To lowest order,
the expansion includes
the diagrams
in Fig.~\ref{fig:lowest}.
The mathematical expression corresponding to the
Wightman self-energy is
\begin{eqnarray}
\Pi^{>}(k) &=& \frac{g^{2}}{2}
\int \frac{d^{4}p}{(2\pi)^{4}} \frac{d^{4}q}{(2\pi)^{4}}
G^{}_{21}(p) G^{}_{12}(q) (2\pi)^{4} \delta^{4}(k+p-q)
%\delta^{4}(q-k'-p)
\nonumber \\
&& +{g^{4}\over 2}
\int
\prod_{i=1}^4\frac{d^{4}q_{i}}{(2\pi)^{4}}
\frac{d^{4}p}{(2\pi)^{4}}
(2\pi)^{12}\delta^{4}(k+q_{4}-q_{1})
\delta^{4}(q_{1} - p - q_{2}) \delta^{4}(p+q_{3} - q_{4})
%\delta^{4}(q_{2} - q_{3} - k')
\nonumber \\
&&\times \Big[ G^{}_{12}(q_{1}) G^{}_{22}(q_{2})
G^{}_{21}(q_{3}) G^{}_{11}(q_{4}) G^{}_{21}(p)
+
G^{}_{11}(q_{1}) G^{}_{12}(q_{2})
G^{}_{22}(q_{3}) G^{}_{21}(q_{4}) G^{}_{12}(p) \Big]
\nonumber \\
&& + \text{other} \; \text{cuts} \; \text{of} \; \mathcal{O}(g^{4}) +
\mathcal{O}(g^{6})
\end{eqnarray}
where `other cuts' do not contribute to the 2 to 2 process.

Neglecting the thermal phase space factor inside
$G_{11}(k)$ and $G_{22}(k)$ yields
\begin{eqnarray}
\Pi^{>}(k) &=&
\frac{g^{2}}{2}
\int \frac{d^{4}p}{(2\pi)^{4}} \frac{d^{4}q}{(2\pi)^{4}}
G^{}_{21}(p) G^{}_{12}(q) (2\pi)^{4} \delta^{4}(k+p-q)
%\delta^{4}(q-k'-p)
\nonumber \\
&&
+{g^{4}\over 2}
\int
\prod_{i=1}^4\frac{d^{4}q_{i}}{(2\pi)^{4}}
\frac{d^{4}p}{(2\pi)^{4}}
(2\pi)^{12}\delta^{4}(k+q_{4}-q_{1})
\delta^{4}(q_{1} - p - q_{2}) \delta^{4}(p+q_{3} - q_{4})
%\delta^{4}(q_{2} - q_{3} - k')
\nonumber \\
&&\times \Big[ G^{}_{12}(q_{1}) G^{*}_{ V}(q_{2})
G^{}_{21}(q_{3}) G^{}_{ V}(q_{4}) G^{}_{21}(p)
+
G_{V}(q_{1}) G^{}_{12}(q_{2})
G^{*}_{ V}(q_{3}) G^{}_{21}(q_{4}) G^{}_{12}(p) \Big]
\non
\end{eqnarray}
where we defined
\begin{eqnarray}
G_{V}(k) = \frac{i}{k^{2} - m_{T} + i|\Pi_I^{ V}|}
\end{eqnarray}

Let's consider the first term.
\begin{eqnarray}
\Pi^>_{(1)}
=
\frac{g^{2}}{2}
\int \frac{d^{4}p}{(2\pi)^{4}} \frac{d^{4}q}{(2\pi)^{4}}
G^{}_{21}(p) G^{}_{12}(q) (2\pi)^{4} \delta^{4}(k+p-q)
\end{eqnarray}
Using the fact that the spectral density is odd
($\rho(-p) = -\rho(p)$)
and by making a suitable change of variable,
this becomes
\begin{eqnarray}
\label{eq:appendixPi1}
\Pi^{>}_{(1)}(k) &=& \frac{g^{2}}{2} \int
\frac{d^{4}p}{(2\pi)^{4}}
\frac{d^{4}q}{(2\pi)^{4}}
\theta(p^{0})\rho(p)
\theta(q^{0})\rho(q) (2\pi)^{4} \nonumber \\
&& {}
\times \Big\{ \left[1+n(|p^{0}|) \right] \left[1+n(|q^{0}|) \right]
\delta^{4}(k-p-q)
\nonumber \\
&& \qquad {}
+ \left[1+n(|p^{0}|) \right]\, n(|q^{0}|)
\delta^{4}(k-p+q) \nonumber \\
&& \qquad {}
+ n(|p^{0}|) \, \left[1+n(|q^{0}|) \right]
\delta^{4}(k+p-q) \nonumber \\
&& \qquad {}
+ n(|p^{0}|)\,  n(|q^{0}|)
\delta^{4}(k+p+q) \Big\}
\end{eqnarray}
Each terms corresponds to a lowest order transition amplitude between
off-shell states coming from the skeleton expansion as defined
previously in $\calK_A$.

The second term is
\begin{eqnarray}
\Pi^{>}_{(2)}(k) &=&
{g^{4}\over 2}
\int
\prod_{i=1}^4\frac{d^{4}q_{i}}{(2\pi)^{4}}
\frac{d^{4}p}{(2\pi)^{4}}
(2\pi)^{12}\delta^{4}(k+q_{4}-q_{1})
\delta^{4}(q_{1} - p - q_{2}) \delta^{4}(p+q_{3} - q_{4})
%\delta^{4}(q_{2} - q_{3} - k')
\nonumber \\
&&\times \Big[ G^{}_{12}(q_{1}) G^{*}_{ V}(q_{2})
G^{}_{21}(q_{3}) G^{}_{ V}(q_{4}) G^{}_{21}(p)
+
G_{V}(q_{1}) G^{}_{12}(q_{2})
G^{*}_{ V}(q_{3}) G^{}_{21}(q_{4}) G^{}_{12}(p) \Big]
\non
\end{eqnarray}
We can use the properties of the spectral function and the explicit
expression of the cut propagators to write the preceding equation as
\begin{eqnarray}
\Pi^{>}_{(2)}(k) &=&
{g^{4}\over 2}
\int \frac{d^{4}q_{1}}{(2\pi)^{4}}
\frac{d^{4}q_{2}}{(2\pi)^{4}}
\frac{d^{4}q_{3}}{(2\pi)^{4}}
(2\pi)^{4}\delta^{4}(k+q_{1}-q_{2} - q_{3})
\nonumber \\
&& \times \theta(q_{1}^{0}) \theta(q_{2}^{0}) \theta(q_{3}^{0})
\rho(q_{1}) \rho(q_{2}) \rho(q_{3}) \,
n(|q_{1}^{0}|)
\left[ 1+n(|q_{2}^{0}|)\right] \left[1+ n(|q_{3}^{0}|)\right] \nonumber
\\
&&\times \Big[ G_{V}^{*}(q_{3} - q_{1}) G_{V}(q_{2} - q_{1}) +
G_{V}^{*}(q_{3} + q_{2}) G_{V}(q_{2} - q_{1}) + G_{V}^{*}(q_{3} -
q_{1}) G_{V}(q_{2} + q_{3})  \nonumber \\
&& \quad {}
+ G_{V}^{*}(q_{2} - q_{1}) G_{V}(q_{3} - q_{1}) +
G_{V}^{*}(q_{2} - q_{1}) G_{V}(q_{2} + q_{3}) + G_{V}^{*}(q_{3} +
q_{2}) G_{V}(q_{3} - q_{1}) \Big] \nonumber \\
&& \quad {} + (\hbox{other processes})
\end{eqnarray}
Each term corresponds to a different channel of a $2 \leftrightarrow 2$
scattering process.
The neglected terms lead to
other processes like $1 \leftrightarrow 3$.
If we now make the lowest order quasi-particle ansatz,
$\rho(p) \to (2\pi)\delta (p^{2} - m^{2}_{T})$,
the above expression can be written as
\begin{eqnarray}
\Pi^{>}_{(2)}(k) &=&
{1\over 2} \int
\frac{d^{3}q_{1}}{(2\pi)^{3}3 E_{1}}
\frac{d^{3}q_{2}}{(2\pi)^{3}3 E_{3}}
\frac{d^{3}q_{3}}{(2\pi)^{3}3 E_{3}}\,
(2\pi)^{4}\delta^{4}(k+q_{1}-q_{2} - q_{3})
\non & & {} \times
n(E_{1})\,
\left[ 1+n(E_{2})\right] \left[1+ n(E_{3})\right] \nonumber \\
&&\times \Big[ M_t^{*} M_u + M_s^{*}M_u
+ M_t^{*}M_s + M_u^{*}M_t + M_u^{*} M_s + M_s^{*} M_t
\Big] \nonumber \\
&& + (\hbox{other processes})
\end{eqnarray}
Here $M_s, M_t, M_u$ represent the scattering amplitude
in the $s, t. u$ channel, respectively.
The terms in the square bracket
is not yet the full scattering element squared since the square terms
are
missing.  We only have the interference terms.
The squared terms should come from iterating Eq.(\ref{eq:appendixPi1}).
We can expand the
spectral density for this one-loop diagram
using Eq.(\ref{eq:Spectral_density_quasi}) and keep only the
relevant terms for $2 \rightarrow 2$ process at lowest order.
This yields, in addition to the pure one-loop contribution,
\begin{eqnarray}
\Pi^{>}_{(1-2)}(k) &=&
\frac{g^{2}}{2}
\int \frac{d^{4}p}{(2\pi)^{4}}
\frac{d^{4}q}{(2\pi)^{4}}
\theta(p^{0}) \theta(q^{0}) (2\pi)^{4} \nonumber \\
&& \times \left[ 2(2\pi)\delta(p^{2} - m_{T}^{2})G_{V}^{*} (q)
\Pi_{I}(q) G_{V} (q) + 2(2\pi)\delta(q^{2} - m_{T}^{2})G_{V}^{*} (p)
\Pi_{I}(p) G_{V} (p)\right] \nonumber \\
&& \times \Big\{
\left[1+n(|p^{0}|) \right] \left[1+n(|q^{0}|) \right]
\delta^{4}(k-p-q)
+ 2\left[1+n(|p^{0}|) \right]\, n(|q^{0}|) 
\delta^{4}(k-p+q)
\non & & {} \quad
+  n(|p^{0}|) \, n(|q^{0}|) 
\delta^{4}(k+p+q) \Big\}
\nonumber \\
&& + (\hbox{other processes})
\end{eqnarray}
Upon using Eqs.(\ref{eq:eq_rels1}) and (\ref{eq:eq_rels2}) this becomes
\begin{eqnarray}
\Pi^{>}_{(1)}(k) &=&
{g^{2}\over 2}
\int \frac{d^{4}p}{(2\pi)^{4}}
\frac{d^{4}q}{(2\pi)^{4}}
\theta(p^{0}) \theta(q^{0}) (2\pi)^{4} \nonumber \\
&& \times \Big\{ \left[1+n(|p^{0}|) \right] (2\pi)\delta(p^{2} -
m_{T}^{2})G_{V}^{*} (q) \Pi^{>}(q) G_{V} (q)\delta^{4}(k-p-q)
\nonumber \\
&& \quad {} + \left[1+n(|p^{0}|) \right] (2\pi)\delta(p^{2} -
m_{T}^{2})G_{V}^{*} (q) \Pi^{<}(q) G_{V} (q)\delta^{4}(k-p+q)
\nonumber \\
&& \quad {}
+  n(|p^{0}|)  (2\pi)\delta(p^{2} - m_{T}^{2})G_{V}^{*}
(q) \Pi^{>}(q) G_{V} (q)  \delta^{4}(k-q+p)
\nonumber \\
&& \quad {}
+   n(|p^{0}|)  (2\pi)\delta(p^{2} -
m_{T}^{2})G_{V}^{*} q) \Pi^{<}(q) G_{V} (q)\delta^{4}(k+p+q)
\Big\}
\nonumber \\
&& {} + (\hbox{other processes})
\label{eq:Pi1}
\end{eqnarray}
The last equation is a self-consistent equation for the self-energy. To
solve it, we can use an iteration procedure. Since we are considering
only the $2 \rightarrow 2$ processes to the lowest order, this means
in the right hand side we should replace
$\Pi^{>,<}(q)$ with the purely one-loop result
\begin{eqnarray}
\Pi^{>}_{(1-1)}(k)
&=&
\frac{g^{2}}{2}
\int \frac{d^{4}p}{(2\pi)^{4}}
\frac{d^{4}q}{(2\pi)^{4}}
\theta(p^{0}) \theta(q^{0}) (2\pi)^{6}
\delta(p^{2} - m_{T}^{2})\delta(q^{2} - m_{T}^{2})\nonumber \\
&& \times \Big\{ 
\left[1+n(|p^{0}|) \right] \left[1+n(|q^{0}|) \right]
\delta^{4}(k-p-q)  \nonumber \\
&& \quad {}
+ 2 \left[1+n(|p^{0}|) \right] \, n(|q^{0}|) 
\delta^{4}(k-p+q) \nonumber \\
&& \quad {}
+  n(|p^{0}|) \, n(|q^{0}|) 
\delta^{4}(k+p+q) \Big\}
\end{eqnarray}
There is an analogous equation for $\Pi^{<}_{(1-1)}(k)$.
Upon substitution, Eq.(\ref{eq:Pi1}) can be shown to contain
the missing square terms
\begin{eqnarray}
\Pi^{>}_{(1)}(k) &=&
\frac{1}{2}
\int
\frac{d^{3}q_{1}}{(2\pi)^{3}2 E_{1}}
\frac{d^{3}q_{2}}{(2\pi)^{3}2 E_{2}}
\frac{d^{3}q_{3}}{(2\pi)^{3}2 E_{3}}
(2\pi)^{4}\delta^{4}(k+q_{1}-q_{2} - q_{3})
\non & & {} \times
 n(E_{1})\,
\left[ 1+n(E_{2})\right] \left[1+ n(E_{3})\right] \,
\Big[ M_u^{*}M_u + M_s^{*}M_s + M_t^{*}M_t  \Big] \nonumber \\
&& + (\hbox{other processes})
\end{eqnarray}
If we combine our two answers, we get
\begin{eqnarray}
\Pi^{>}(k) &=& \Pi^{>}_{(1)}(k) + \Pi^{>}_{(2)}(k) \nonumber \\
& = &
\frac{1}{2}
\int
\frac{d^{3}q_{1}}{(2\pi)^{3}2 E_{1}}
\frac{d^{3}q_{2}}{(2\pi)^{3}2 E_{2}}
\frac{d^{3}q_{3}}{(2\pi)^{3}2 E_{3}}
(2\pi)^{4}\delta^{4}(k+q_{1}-q_{2} - q_{3})
\non & & {} \times
 n(E_{1})\,
\left[ 1+n(E_{2})\right] \left[1+ n(E_{3})\right] \,
\Big| M_u + M_s + M_t \Big|^2
\end{eqnarray}
This is in accordance with our general result expressed in
equation (\ref{eq:PiBoltz}).


\begin{thebibliography}{99}


\bibitem{McLerran_etal_1994_1}
L. D. McLerran and R. Venugopalan, Phys. Rev. D {\bf 49}, 2233 (1994)
[arXiv:hep-ph/9309289].

\bibitem{McLerran_etal_1994_2}
L. D. McLerran and R. Venugopalan, Phys. Rev. D {\bf 49}, 3352 (1994)
[arXiv:hep-ph/9311205].

\bibitem{McLerran_2005}
L. McLerran, Nucl. Phys. A {\bf 752}, 355 (2005).

\bibitem{Venugopalan_2005}
R. Venugopalan, Eur. Phys. J. C {\bf 43}, 337 (2005)
[arXiv:hep-ph/0502190].

\bibitem{Kharzeev:2000ph}
  D.~Kharzeev and M.~Nardi,
  %``Hadron production in nuclear collisions at RHIC and high density
%%QCD,''
  Phys.\ Lett.\ B {\bf 507}, 121 (2001)
  [arXiv:nucl-th/0012025].
  %%CITATION = NUCL-TH 0012025;%%

\bibitem{Kharzeev:2001yq}
  D.~Kharzeev, E.~Levin and M.~Nardi,
  %``The onset of classical QCD dynamics in relativistic heavy ion
  %collisions,''
  Phys.\ Rev.\ C {\bf 71}, 054903 (2005)
  [arXiv:hep-ph/0111315].
  %%CITATION = HEP-PH 0111315;%%

\bibitem{Kharzeev:2002ei}
  D.~Kharzeev, E.~Levin and M.~Nardi,
  %``QCD saturation and deuteron nucleus collisions,''
  Nucl.\ Phys.\ A {\bf 730}, 448 (2004)
  [Erratum-ibid.\ A {\bf 743}, 329 (2004)]
  [arXiv:hep-ph/0212316].
  %%CITATION = HEP-PH 0212316;%%

\bibitem{Schwinger_1961}
J. Schwinger, J. Math. Phys. {\bf 2}, 407 (1961).

\bibitem{Keldysh_1964}
L. V. Keldysh, Sov. Phys. JETP {\bf 20}, 1018 (1964).

\bibitem{Jeon_1995}
S. Jeon, Phys. Rev. D {\bf 52}, 3591 (1995) [arXiv: hep-ph/9409250].

\bibitem{Jeon_Yaffe_1996}
S. Jeon and L. G. Yaffe, Phys. Rev. D {\bf 53}, 5799 (1996) [arXiv:
hep-ph/9512263].

\bibitem{AMY_2000}
P. Arnold, G. D. Moore and L. G. Yaffe, JHEP {\bf 0011}, 001 (2000)
[arXiv:
hep-ph/0010177].

\bibitem{AMY_2003}
P. Arnold, G. D. Moore and L. G. Yaffe, JHEP {\bf 0305}, 051 (2003)
[arXiv:
hep-ph/0302165].

\bibitem{AMY_2003_2}
P. Arnold, G. D. Moore and L. G. Yaffe, JHEP {\bf 0301}, 030 (2003)
[arXiv:
hep-ph/0209353].

\bibitem{Kadanoff_Martin_1963}
L. P. Kadanoff and P. C. Martin, Ann. Phys. NY {\bf 24}, 419 (1963).

\bibitem{Carrington_etal_2005}
M. E. Carrington and S. Mr\'{o}wczy\'{n}ski, Phys. Rev. D {\bf 71}
065007 (2005) [arXiv: hep-ph/0406097].

\bibitem{Mrowczynski:1989bu}
S.~Mrowczynski and P.~Danielewicz,
%``Green Function Approach To Transport Theory Of Scalar Fields,''
Nucl.\ Phys.\ B {\bf 342}, 345 (1990).
%%CITATION = NUPHA,B342,345;%%

\bibitem{Baier_etal_2001}
R. Baier {\it et al.}, Phys. Lett. B {\bf 502}, 51 (2001) [arXiv:
hep-ph/0009237].

\bibitem{Srivastava_Geiger_1999}
D. K. Srivastava and K. Geiger, Nucl. Phys. A {\bf 647}, 136 (1999)
[arXiv:
nucl-th/9806050].

\bibitem{Wong_2004}
S. M. H. Wong, arXiv: hep-ph/0404222.

\bibitem{Xu_Greiner_2005}
Z. Xu and C. Greiner,
Phys.\ Rev.\ C {\bf 71}, 064901 (2005) [arXiv:hep-ph/0406278].

\bibitem{Jeon_Ellis_1998}
S. Jeon and P. J. Ellis, Phys. Rev. D {\bf 58}, 045013 (1998) [arXiv:
hep-ph/9802246].

\bibitem{Gagnon_etal_2005}
J.-S. Gagnon, F. Fillion-Gourdeault and S. Jeon, Quark Matter 2005
proceedings, arXiv: hep-ph/0510367.

\bibitem{Weinstock_2005}
S.~Weinstock,
%``Boltzmann collision term,''
Phys. Rev. D {\bf 73}, 025005 (2006) [arXiv:hep-ph/0510417].

\bibitem{Elze_Heinz_1989}
H. T. Elze and U. W. Heinz,
Phys.\ Rept.\  {\bf 183}, 81 (1989).

\bibitem{Mrowczynski_1997}
S. Mrowczynski, Phys. Rev. D {\bf 56}, 2265 (1997).
[arXiv:hep-th/9702022].

\bibitem{Baym_Kadanoff_1962}
L.P. Kadanoff and G. Baym, {\it Quantum Statistical Mechanics}
(Benjamin, New
York (1962)).

\bibitem{Groot_etal_1980}
S.R. de Groot, W.A. van Leeuwen and Ch.G. van Weert, {\it Relativistic
Kinetic
Theory} (North-Holland, Amsterdam (1980)).

\bibitem{Kobes_Semenoff_1985}
R. L. Kobes and G. W. Semenoff, Nucl. Phys. B {\bf 260}, 714 (1985);
Nucl.
Phys. B {\bf 272}, 329 (1986).

\bibitem{Diagrammar}
G. 't Hooft and M. Veltman, ``Diagrammar'', CERN Yellow Report 73-9
(1973).

\bibitem{Weinberg}
S. Weinberg, {\it The Quantum Theory of Fields} (Cambridge University
Press, Cambridge, (2002)).

\bibitem{Landsman_VanWeert_1987}
N. P. Landsman and Ch. G. van Weert, Phys. Rep. {\bf 145}, 141 (1987).

\bibitem{Landsman:1988ta}
 N.~P.~Landsman,
%``Nonshell Unstable Particles In Thermal Field Theory,''
Annals Phys.\  {\bf 186}, 141 (1988).
%%CITATION = APNYA,186,141;%%

\bibitem{Altherr_Seibert_1994}
T. Altherr and D. Seibert, Phys. Lett. B {\bf 333}, 149 (1994) [arXiv:
hep-ph/9405396].

\bibitem{Altherr_1995}
T. Altherr, Phys. Lett. B {\bf 341}, 325 (1995) [arXiv:
hep-ph/9407249].

\bibitem{Bedaque_1995}
P. F. Bedaque,  Phys. Lett. B {\bf 344}, 23 (1995) [arXiv:
hep-ph/9410415].

\bibitem{Greiner_Leupold_1998}
C. Greiner and S. Leupold, Annals Phys. {\bf 270}, 328 (1998) [arXiv:
hep-ph/9802312].

\bibitem{Greiner_Leupold_1999}
C. Greiner and S. Leupold, Eur. Phys. J. C {\bf 8}, 517 (1999) [arXiv:
hep-ph/9804239].

\bibitem{Braaten_Pisarski_1990}
E.~Braaten and R.~D.~Pisarski,
%``Soft Amplitudes In Hot Gauge Theories: A General Analysis,''
Nucl.\ Phys.\ B {\bf 337}, 569 (1990).

\bibitem{Frenkel_Taylor_1990}
J.~Frenkel and J.~C.~Taylor,
%``High Temperature Limit Of Thermal QCD,''
Nucl.\ Phys.\ B {\bf 334}, 199 (1990).

%\bibitem{Boyanovsky_etal_2000}
%D. Boyanovsky, H. H. de Vega and S.-Y. Wang, Phys. Rev. D, {\bf 61},
%%065006
%(2000) [arXiv: hep-ph/9909369].

\bibitem{Aarts_Berges_2001}
G.~Aarts and J.~Berges,
%``Nonequilibrium time evolution of the spectral function in quantum field
%theory,''
Phys. Rev. D {\bf 64}, 105010 (2001) [arXiv:hep-ph/0103049].

\bibitem{Majumder_Gale_2002}
  A.~Majumder and C.~Gale,
  %``On the imaginary parts and infrared divergences of two-loop vector  boson
  %self-energies in thermal QCD,''
  Phys.\ Rev.\ C {\bf 65}, 055203 (2002)
  [arXiv:hep-ph/0111181].
  %%CITATION = HEP-PH 0111181;%%

\bibitem{Gale:2005ri}
  C.~Gale,
  %``In-medium effects on electromagnetic probes,''
  Eur.\ Phys.\ J.\ C {\bf 43}, 381 (2005)
  [arXiv:hep-ph/0504103].
  %%CITATION = HEP-PH 0504103;%%

\bibitem{Mallik:2001qa}
  S.~Mallik,
  %``Scattering amplitude and shift in self-energy in medium,''
  Eur.\ Phys.\ J.\ C {\bf 24}, 143 (2002)
  [arXiv:hep-th/0108139].
  %%CITATION = HEP-TH 0108139;%%

\bibitem{Eletsky:2001bb}
  V.~L.~Eletsky, M.~Belkacem, P.~J.~Ellis and J.~I.~Kapusta,
  %``Properties of rho and omega mesons at finite temperature and density as
  %inferred from experiment,''
  Phys.\ Rev.\ C {\bf 64}, 035202 (2001)
  [arXiv:nucl-th/0104029].
  %%CITATION = NUCL-TH 0104029;%%

\bibitem{Kapusta:2001jw}
  J.~I.~Kapusta and S.~M.~H.~Wong,
  %``Two-loop self-energy and multiple scattering at finite temperature,''
  Phys.\ Rev.\ D {\bf 64}, 045008 (2001)
  [arXiv:hep-th/0103065].
  %%CITATION = HEP-TH 0103065;%%

\bibitem{Amelino-Camelia:1999yr}
  G.~Amelino-Camelia and J.~I.~Kapusta,
  %``Neutral kaon system in dense matter and heavy-ion collisions,''
  Phys.\ Lett.\ B {\bf 465}, 291 (1999)
  [arXiv:hep-ph/9907508].
  %%CITATION = HEP-PH 9907508;%%

\bibitem{Gao:1998mn}
  S.~Gao, C.~Gale, C.~Ernst, H.~Stoecker and W.~Greiner,
  %``rho meson broadening in hot and dense hadronic matter,''
  arXiv:nucl-th/9812059.
  %%CITATION = NUCL-TH 9812059;%%

\bibitem{Eletsky:1998jm}
  V.~L.~Eletsky and J.~I.~Kapusta,
  %``Dispersion relation of the rho meson at finite temperature and
  %density,''
  arXiv:nucl-th/9810052.
  %%CITATION = NUCL-TH 9810052;%%

\bibitem{Leupold:1999ga}
  S.~Leupold,
  %``Towards a test particle description of transport processes for states
  %with
  %continuous mass spectra,''
  Nucl.\ Phys.\ A {\bf 672}, 475 (2000)
  [arXiv:nucl-th/9909080].
  %%CITATION = NUCL-TH 9909080;%%

\bibitem{Ivanov:1999tj}
  Y.~B.~Ivanov, J.~Knoll and D.~N.~Voskresensky,
  %``Resonance Transport and Kinetic Entropy,''
  Nucl.\ Phys.\ A {\bf 672}, 313 (2000)
  [arXiv:nucl-th/9905028].
  %%CITATION = NUCL-TH 9905028;%%




\bibitem{Juchem:2004cs}
  S.~Juchem, W.~Cassing and C.~Greiner,
  %``Nonequilibrium quantum-field dynamics and off-shell transport for
  %phi**4-theory in 2+1 dimensions,''
  Nucl.\ Phys.\ A {\bf 743}, 92 (2004)
  [arXiv:nucl-th/0401046].
  %%CITATION = NUCL-TH 0401046;%%

\bibitem{Cassing:2000ch}
  W.~Cassing and S.~Juchem,
  %``Equilibration within a semiclassical off-shell transport approach,''
  Nucl.\ Phys.\ A {\bf 677}, 445 (2000)
  [arXiv:nucl-th/0003002].
  %%CITATION = NUCL-TH 0003002;%%

\bibitem{Lindner:2005kv}
  M.~Lindner and M.~M.~Muller,
  %``Comparison of Boltzmann equations with quantum dynamics for scalar
  %fields,''
  arXiv:hep-ph/0512147.
  
\bibitem{Knoll:2001jx}
  J.~Knoll, Y.~B.~Ivanov and D.~N.~Voskresensky,
  %``Exact Conservation Laws of the Gradient Expanded Kadanoff-Baym Equations,''
  Annals Phys.\  {\bf 293}, 126 (2001)
  [arXiv:nucl-th/0102044].

\bibitem{Ivanov:1998nv}
  Y.~B.~Ivanov, J.~Knoll and D.~N.~Voskresensky,
  %``Self-consistent approximations to non-equilibrium many-body theory,''
  Nucl.\ Phys.\ A {\bf 657}, 413 (1999)
  [arXiv:hep-ph/9807351].


\bibitem{Gelis_Venugopalan_2006}
F. Gelis and R. Venugopalan, arXiv: hep-ph/0601209.


\end{thebibliography}
\end{document}